\documentstyle[12pt]{article}

\input epsf

\newcommand{\beq}{\begin{equation}}
\newcommand{\eeq}{\end{equation}}
\newcommand{\bd}{\begin{displaymath}}
\newcommand{\ed}{\end{displaymath}}
\newcommand{\bea}{\begin{eqnarray}}
\newcommand{\eea}{\end{eqnarray}}
\newcommand{\nn}{\nonumber}
\newcommand{\la}{\langle}
\newcommand{\ra}{\rangle}

\font\mybbs=msbm10 at 9pt
\def\bbs#1{\hbox{\mybbs#1}}
\font\mybb=msbm10 at 12pt
\def\bb#1{\hbox{\mybb#1}}

\def\appendix#1{
  \addtocounter{section}{1}
  \setcounter{equation}{0}
  \renewcommand{\thesection}{\Alph{section}}
  \section*{Appendix \thesection\protect\indent \parbox[t]{11.715cm} {#1}}
  \addcontentsline{toc}{section}{Appendix \thesection\ \ \ #1}}

\def\IC{{\bb C}}
\def\IR{{\bb R}}
\def\IZ{{\bb Z}}

\def\IRs{{\bbs R}}

\def\tr{{\rm tr}}
\def\ps2{{\bar{\psi}\psi}}
\def\e{{\rm e}}
\def\P{{\bf P}}
\def\Q{{\bf Q}}
\def\L{\overline{\bf P}}
\def\M{\overline{\bf Q}}
\def\Tr{{\rm Tr}}

\makeatletter
\newdimen\normalarrayskip              
\newdimen\minarrayskip                 
\normalarrayskip\baselineskip
\minarrayskip\jot
\newif\ifold             \oldtrue            \def\new{\oldfalse}
\def\arraymode{\ifold\relax\else\displaystyle\fi} 
\def\@arrayskip{\ifold\baselineskip\z@\lineskip\z@
     \else
     \baselineskip\minarrayskip\lineskip2\minarrayskip\fi}
\def\@arrayclassz{\ifcase \@lastchclass \@acolampacol \or
\@ampacol \or \or \or \@addamp \or
   \@acolampacol \or \@firstampfalse \@acol \fi
\edef\@preamble{\@preamble
  \ifcase \@chnum
     \hfil$\relax\arraymode\@sharp$\hfil
     \or $\relax\arraymode\@sharp$\hfil
     \or \hfil$\relax\arraymode\@sharp$\fi}}
\def\@array[#1]#2{\setbox\@arstrutbox=\hbox{\vrule
     height\arraystretch \ht\strutbox
     depth\arraystretch \dp\strutbox
     width\z@}\@mkpream{#2}\edef\@preamble{\halign \noexpand\@halignto
\bgroup \tabskip\z@ \@arstrut \@preamble \tabskip\z@ \cr}%
\let\@startpbox\@@startpbox \let\@endpbox\@@endpbox
  \if #1t\vtop \else \if#1b\vbox \else \vcenter \fi\fi
  \bgroup \let\par\relax
  \let\@sharp##\let\protect\relax
  \@arrayskip\@preamble}
\makeatother

\newlength{\extraspace}
\newlength{\extraspaces}
\begin{document}

\renewcommand{\footnotesize}{\small}

\addtolength{\baselineskip}{.8mm}

\thispagestyle{empty}

\begin{flushright}
\baselineskip=12pt
PUPT-1931\\
NBI-HE-00-23\\
\end{flushright}
\vspace{1.2cm}

\begin{center}

\baselineskip=24pt

{\Large\bf Fermionic Quantum Gravity}\\
\medskip
\vskip0.4in

\baselineskip=12pt

{\bf Lori D. Paniak}
\smallskip
\medskip

{\it Joseph Henry Laboratories \\ Princeton
University\\
Princeton, New Jersey 08544, U.S.A.}\\{\tt paniak@feynman.princeton.edu}

\vskip 0.3in

{\bf Richard J. Szabo}
\smallskip
\medskip

{\it The Niels Bohr Institute\\ Blegdamsvej 17, DK-2100 Copenhagen \O,
Denmark}\\{\tt szabo@nbi.dk}

\vskip1in

{\bf ABSTRACT}

\begin{center}
\begin{minipage}{15cm}

We study the statistical mechanics of random surfaces generated by $N\times N$
one-matrix integrals over anti-commuting variables. These
Grassmann-valued matrix models are shown to be equivalent to 
$N\times N$ unitary versions of generalized Penner matrix models. 
We explicitly solve for the combinatorics
of 't~Hooft diagrams of the matrix integral and develop an orthogonal
polynomial formulation of the statistical theory. An examination of 
the large $N$
and double scaling limits of the theory shows that the genus expansion is a
Borel summable alternating series which otherwise coincides with
two-dimensional quantum gravity in the continuum limit. We demonstrate that the
partition functions of these matrix models belong to the relativistic Toda
chain integrable hierarchy. The corresponding string equations and Virasoro
constraints are derived and used to analyse the generalized KdV flow structure
of the continuum limit.

\end{minipage}
\end{center}

\end{center}

\vfill
\newpage

\pagestyle{plain}
\setcounter{page}{1}
\stepcounter{subsection}

\tableofcontents

\section{Introduction and Summary}
\setcounter{equation}{0}

Matrix models \cite{mehta} provide a useful framework in which to extract
quantitative information about complex statistical systems given only their
general symmetry properties. The key feature which enables this description is
the existence of universal behaviour in these models. Over the years random
matrix models have been utilized in a wide range of studies in condensed matter
systems \cite{condrev}, quantum chromodynamics \cite{qcdrev}, and
low-dimensional string theory \cite{fgz,2Dbook}. They are also useful for
solving various statistical mechanical models and combinatorial problems
\cite{matrixcomb}. One interpretation of random matrix ensembles is in terms of
generators of 't~Hooft diagram expansions which correspond to discretizations
of compact Riemann surfaces. The continuum limits of such discretizations
are realized as phase transitions in the statistical mechanics of the
matrix model and characterized by universal data at the critical points.

Despite the success of matrix model technology over the years, there remain
several issues in random surface theory which are not adequately addressed
by standard approaches. The simplest, and one of the most fundamental issues
regards a self-consistent definition of the vacuum of pure two-dimensional
quantum gravity. The conventional matrix model methods for investigating this
theory lead to inconsistencies. For instance, bosonic matrix integrals are
typically ill-defined in the physically interesting region of parameter space
corresponding to the non-perturbative continuum limit. Furthermore, they are
unable to describe the coupling of gravity to conformal matter fields of
central charge $c>1$. It has been suggested \cite{BP} that above this $c=1$
barrier the two-dimensional geometry degenerates into a tree-like or branched
polymer phase.

In this paper we will study a model of discretized random surfaces which is in
part motivated by such issues. It is parameterized by a zero-dimensional action
involving matrices with Grassmann-valued elements \cite{mz}. This model is
known as the adjoint fermion one-matrix model and it is defined by the
partition function
\beq
Z_N=\int\limits_{{\rm Gr}(N)^c}d\psi~d\bar\psi~\e^{N\,\tr\,V(\ps2)}
\label{1mat}\eeq
where
\beq
V(\ps2)=\sum_{k=1}^K\frac{g_k}{k}\,\left(\ps2\right)^k
\label{pot}\eeq
is a polynomial potential of order $K\leq\infty$, and
$d\psi~d\bar\psi=\prod_{i,j}d\psi_{ij}~d\bar\psi_{ij}$ is the standard Berezin
measure on
the complexified Grassmann algebra ${\rm Gr}(N)^c={\rm Gr}(N)\otimes_\IRs\IC$.
For any matrix $X$ the trace is normalized as $\tr\,X=\sum_iX_{ii}$. The matrix
elements $\psi_{ij}$ and $\bar\psi_{ij}$, $i,j=1,\dots,N$, are independent,
complex-valued Grassmann numbers with the usual rules for complex conjugation,
$\bar\psi^*_{ij}=-\psi_{ij}$, $\psi_{ij}^*=-\bar\psi_{ij}$, and
$(\bar\psi_{ij}\psi_{kl})^*=\psi_{kl}^*\bar\psi_{ij}^*$ (This ensures that the
generating function (\ref{1mat}) is real-valued). The large $N$ limit of this
model has been studied in \cite{mz}--\cite{ss1} and shown to exhibit 
the range of critical behaviour seen in the usual bosonic one-matrix
models. The topological expansion of the matrix integral (\ref{1mat}) was
considered in \cite{acm} and some novel critical behaviour was observed in
\cite{mss}. The fermionic matrix model can also be coupled to an ordinary
complex bosonic one to produce a supersymmetric matrix model, which has been
used to describe solutions of certain combinatorial problems \cite{susy} and
also to generate statistical models of branched polymers \cite{amzpoly}.

The most attractive feature of the fermionic matrix model (\ref{1mat}) is that
the integration over Grassmann variables is always well-defined and the theory
yields finite,  computable observables. The dimension $N$ of the
Grassmann matrices acts as a cutoff on the number of terms in the partition
function (\ref{1mat}) which counts all fat-graphs generated. The fact that the
partition function is a polynomial in the coupling constants $g_k$ for finite
$N$, coupled with the fact that inner products in the Grassmann integration
measure are effectively calculable, leads to explicit expressions
for statistical quantities in this class of models. In the following we will
use this property to solve analytically the combinatorial problem of
counting the dynamical triangulations generated by the fermionic partition
function via an explicit determinant representation of (\ref{1mat}).  

Since the generating function (\ref{1mat}) is always a well-defined convergent
object at finite $N$, one might expect that its large $N$ limit is also
well-defined. In perturbation theory, the Feynman rules for 
fermionic matrices include a factor of $-1$ for each fermion loop
thereby resulting in cancellations between large numbers of Feynman diagrams.
This observation has
led to the conjecture that the genus expansion of the matrix integral
(\ref{1mat}) in the large $N$ limit is an alternating series which may
be Borel summable. The evidence for this striking property of the
adjoint fermion one-matrix model is its intimate relationship with generalized
Penner matrix models \cite{mz,acm,ss1} which are defined by the Hermitian
matrix integrals
\beq
Z_{N,\alpha}^{\rm H}=
\int\limits_{u(N)}d\phi~\e^{N\,\tr\bigl(V(\phi)+\alpha\log\phi\bigr)}
\label{ZNH}\eeq
where $d\phi=\prod_{i,j}d\phi_{ij}$ is the Haar measure on the Lie algebra
$u(N)$ of the $N\times N$ unitary group $U(N)$. It has been argued that the
matrix model (\ref{ZNH}) is equivalent to (\ref{1mat}) for $\alpha=-2$, with
the Hermitian matrix $\phi$ identified with $\ps2$. This has been established
in the large $N$ limit by a saddle-point computation on $\phi\sim\ps2$
\cite{ss1}, and by showing that their Schwinger-Dyson (loop) equations are
identical order by order in the large $N$ expansions of the matrix integrals
\cite{mz,acm}. However, for $\alpha<0$ the integration over Hermitian matrices
in (\ref{ZNH}) is not well-defined because of the logarithmic divergence at
$\phi=0$. In this region the integral can only be defined by analytical
continuation and the matrix model only makes sense at $N=\infty$. This
continuation produces complex-valued endpoints for the support of the
corresponding distribution of $\phi$ eigenvalues, analogous to what 
naturally occurs in fermionic matrix models
\cite{mz,ss1}. It is this complexification of the support of the
distribution of eigenvalues that is asserted to result in an alternating
genus expansion. Moreover, it can be argued \cite{acm} that the topological
expansion coincides, modulo these sign factors, with the usual Painlev\'e
expansions of Hermitian matrix integrals with polynomial potentials \cite{fgz}.

The near equivalence of the genus expansions strongly suggests a correspondence
between the Feynman graph expansions in the bosonic and fermionic models. One
way of understanding these issues is to appeal to a simplified version of the
matrix integral (\ref{1mat}), namely a fermionic vector model defined by
anticommuting vector elements $\psi_i$ and $\bar\psi_i$, $i=1,\dots,N$
\cite{ss2}. A fermionic vector model is related to a bosonic vector model,
defined with the same polynomial potential $V$, by a simple mapping of its
Feynman diagrams and an analytical continuation $N\mapsto-\frac N2$ of the
vector dimension which results in a bosonic perturbation series with a cutoff
on the number of terms generated. This analytical continuation can be
understood in terms of a complex-valued saddle-point of the bosonic model in
the large $N$ limit, leading to an alternating ``genus expansion'' which
produces a well-defined, unique function \cite{ss2}. As bosonic vector models
generate random models of branched polymers, the fermionic vector model thereby
produces a statistical theory of branched polymers whose continuum
characteristics are equivalent order by order in large $N$ to the conventional
models, but whose double-scaling expansion is a non-perturbatively well-defined
function.

The arguments given in \cite{acm} are based on the class of models with odd
polynomial potential, i.e. $V(-\phi)=-V(\phi)$. In these cases, the endpoints
of the support contour of the spectral density are located symmetrically about
the origin on the imaginary axis in the complex plane \cite{mz,ss1}. Generic
polynomial potentials are difficult to treat because the corresponding support
contours are located asymmetrically in the complex plane and the loop equations
become cumbersome to deal with. Moreover, although the equations of motion for
the matrix models (\ref{1mat}) and (\ref{ZNH}) are identical at each order of
the large $N$ expansion, beyond the leading order they must be solved with
different boundary conditions and the non-perturbative solutions are not the
same in the two cases \cite{acm}. One of the main results of this paper will be
the clarification of the role played by generalized Penner models in fermionic
matrix integrals. The analytical expressions for the partition function
(\ref{1mat}) at finite $N$ will be interpreted in terms of Toeplitz
determinants, which will naturally lead to a proof of the equivalence of the
fermionic one-matrix model and a {\it unitary} version of the Penner one-matrix
model,
\beq
Z_N^{\rm U}=\int\limits_{U(N)}[dU]~\e^{N\,\tr\,\bigl(V(U)-\log U\bigr)}
\label{ummdef1}\eeq
where $[dU]$ is the invariant Haar measure on the unitary group $U(N)$. The
important feature of this equivalence is that it holds for any matrix dimension
$N$. The finiteness of the fermionic matrix model at finite $N$ is captured by
the compactness of the field variable in (\ref{ummdef1}). As we will see, the
logarithmic term in the effective potential of (\ref{ummdef1}) gives certain
restrictions on the configurations which contribute to the partition function
and its observables when the unitary matrix model is treated from a group
theoretic perspective. These restrictions naturally reproduce the basic
characteristics of the matrix integral (\ref{1mat}), and moreover show
precisely how the analytic, functional forms of the models (\ref{1mat}) and
(\ref{ummdef1}) are equivalent for any value of $N$. We will also prove
directly that the loop equations of the fermionic and unitary matrix models are
identical, and that they admit the same, perturbative Gaussian boundary
conditions. It is a remarkable fact that a unitary matrix model such as
(\ref{ummdef1}) admits an interpretation in terms of random surfaces.

The biggest advantage of the mapping of the fermionic matrix model to a unitary
one is that we now have an eigenvalue model to analyse. Grassmann-valued
matrices are not diagonalizable, and a direct analysis of fermionic matrix
models is only possible using the method of loop equations \cite{mz,ss1}. In
this paper we shall exploit this eigenvalue representation to give a systematic
and detailed account of the exotic properties possessed by fermionic matrix
models. In particular we will develop a generalization of the orthogonal
polynomial technique on the circle which is necessary to study the unitary
matrix model defined by (\ref{ummdef1}), and describe some of their functional
analytic properties. These methods will naturally lead, irrespective of the
parity of the potential $V$, to a precise investigation of the double-scaling
limit of the fermionic matrix model which determines a non-perturbative,
all-genus continuum theory of two-dimensional quantum gravity coupled to
matter. We will show that the fermionic matrix model leads to a Borel summable
genus expansion, in contradistinction to Hermitian one-matrix models. Unlike
the vector models, the diagrammatic, finite $N$ situation is not quite so
simple in the case of fermionic matrices, but the final qualitative result will
be the same. In particular, we shall obtain a non-perturbative definition of
pure two-dimensional quantum gravity which is equivalent to the usual models
order by order in the genus expansion, but which is given by a well-defined
unique generating function. This non-perturbative model is what we shall call
``fermionic quantum gravity'' in the following.

Another advantage of the eigenvalue representation is that it will allow us
describe the fermionic one-matrix model (\ref{1mat}) as an integrable system.
Generally, matrix models, being examples of exactly solvable systems, are
intimately related to certain reductions of well-known integrable hierarchies
of differential equations \cite{inthi}. In the usual bosonic Hermitian
one-matrix models, describing two-dimensional quantum gravity, the partition
functions are related to the $\tau$-functions of the integrable Toda chain
hierarchy. In the following we will see that the fermionic gravity theory is
related to the $\tau$-function of a particular deformation of this hierarchy
known as the relativistic Toda chain. We can therefore understand many of the
novel aspects of the fermionic theory, such as what sort of matter coupling it
involves, in terms of this deformation. Furthermore, we will develop an
operator theoretic approach to the equations of motion of the matrix integral
(\ref{1mat}) and gain in this setting a precise picture of how this deformation
leads to a well-defined Borel summable partition function of the statistical
model. This sets the stage for a description of the integrable differential
hierarchies satisfied by the continuum  theory which generalizes the usual KdV
flow structure of gravity \cite{fgz,2Dbook}. It will also indicate that
the 
partition function (\ref{1mat}) in the continuum limit serves as a concrete,
non-perturbative definition of two-dimensional quantum gravity.

The organization of the remainder of this paper is as follows. In section 2 we
will derive an analytical expression for the partition function (\ref{1mat})
and use it to describe the solution to the problem of counting fermionic
fat-graphs. We shall see that the matrix integral generates a novel class of
discretizations of Riemann surfaces whose polygons are two-colourable. We will
also briefly describe what sort of matter-coupled gravity theory is represented
by the fermionic one-matrix model, and prove its equivalence to the unitary
one-matrix model (\ref{ummdef1}). In section 3 we will present the formal
orthogonal polynomial solution of the fermionic matrix model. In section 4 we
will describe the large $N$ limit of the model from the perspective of these
orthogonal polynomials. We will find that the adjoint fermion one-matrix model
actually possesses a sort of ``internal'' branched polymer critical behaviour
which can be understood in terms of the fermionic characteristics of its
fat-graph combinatorics. We shall also demonstrate that the usual universality
classes of two-dimensional gravity arise, and that they cannot be smoothly
connected to the branched polymer phases of the theory. Of course, the polymer
effect is washed out at $N=\infty$ because the surface effect is of higher
order in $N$. We also construct the genus expansion of the theory and establish
its Borel summability explicitly. In section 5 we present the operator approach
to the fermionic matrix model and use it to interpret the partition function as
a $\tau$-function. We derive the Virasoro constraints satisfied by this
$\tau$-function, which at the same time establishes explicitly the equivalence
of the loop equations of the matrix models (\ref{1mat}), (\ref{ZNH}) with
$\alpha=-2$, and (\ref{ummdef1}), in the large $N$ expansion. We then use these
descriptions to give an functional analytic explanation for the nature of the
topological expansion in the fermionic one-matrix model. Finally, in appendix A
we describe properties of observables of the statistical model (\ref{1mat})
using the unitary matrix model and its orthogonal polynomials, while in
appendix B we show that the orthogonal polynomials of the Gaussian
fermionic one-matrix model are given in terms of confluent hypergeometric
polynomials.

\section{Determinant Representations of the Partition Function}
\setcounter{equation}{0}

In this section we will present an analytic solution to the problem of counting
fermionic fat-graphs. We will obtain a closed expression for the generating
function of such discretized random surfaces, and at the same time we will be
naturally led to the equivalence of the fermionic matrix model and the unitary
matrix model which will occupy the bulk of our analysis in subsequent sections.
We will also obtain in this way a geometric picture of what sort of theory of
random surfaces is represented by the adjoint fermion one-matrix model and also
of the precise role of the corresponding Penner interaction.

\subsection{Hermitian Representation}

One intriguing feature of the fermionic one-matrix model (\ref{1mat}) is the
extent to which it can be solved at finite-$N$. This is a consequence of
the convergence of the integration over Grassmann variables, in contrast to the
case of bosonic matrix models. The combinatorics of the fat-graph expansion of
the fermionic partition function $Z_N$ can be
determined by using the observation of \cite{mss} that the partition function
and observables of the matrix model (\ref{1mat}) can be evaluated explicitly by
introducing two $N\times N$ Hermitian matrices $\phi$ and $\lambda$ defined by
\beq
1=\int\limits_{u(N)}d\phi~\delta(\phi-\ps2)~~~~~~,
{}~~~~~~\delta(\phi)=\int\limits_{u(N)}
\frac{d\lambda}{(2\pi)^{N^2}}~\e^{i\,\tr\,\lambda\phi}
\eeq
By inserting these definitions into
(\ref{1mat}), the fermionic partition function can be written as
\bea
Z_N&=&\int\limits_{u(N)}d\phi~\e^{N\,\tr\,V(\phi)}\int\limits_{{\rm
Gr}(N)^c}d\psi~d\bar\psi
\int\limits_{u(N)}\frac{d\lambda}{(2\pi)^{N^2}}~\e^{i\,\tr\,\lambda(\phi-\ps2)}
\nn\\&=&\frac{1}{(2\pi)^{N^2}}\int\!\!\!\!\!\int\limits_{u(N)}d\phi~
d\lambda~\e^{N\,\tr \,V(\phi)+i\,\tr\,\lambda\phi}~{\det}^N(-i\lambda)
\nn\\&=&\int\limits_{u(N)}d\phi~\e^{N
\,\tr\,V(\phi)}~{\det}^N\left(\mbox{$-\frac{\partial}{\partial\phi}$}
\right)\delta(\phi)
\label{2mat}\eea
so that
\beq
Z_N={\det}^N\left(\mbox{$\frac{\partial}{\partial\phi}$}\right)~\e^{N\,\tr
\,V(\phi)}\Bigm|_{\phi=0}
\label{ZNdet}\eeq
A similar expression can also be derived for the correlators of the fermionic
one-matrix model \cite{mss,ss1}. However, it is difficult
to write down an explicit expression for (\ref{ZNdet}) which is informative and
amenable to analysis.

To evaluate the determinant in (\ref{ZNdet}) explicitly, we shall instead
 write the fermionic matrix model as a random theory of the two Hermitian
matrices $\phi$ and $\lambda$ introduced above, keeping only the second
line of the identity (\ref{2mat}). In this way, the fermionic one-matrix model
can be written as the Hermitian two-matrix model
\beq
Z_N=\frac{N^{N(N+1)}}{(2\pi)^{N^2}}\int\!\!\!\!\!\int\limits_{u(N)}
d\phi~d\lambda~\e^{N\,\tr\bigl(V(\phi)+\log(-i\lambda)+i\lambda\phi\bigr)}
\label{ZN2mat}\eeq
where we have rescaled $\lambda_{ij}\to N\cdot\lambda_{ij}$. The doubling of
degrees of freedom is required to compensate for both the matrix $\psi$ and its
adjoint $\bar\psi$. The Hermitian two-matrix model
(\ref{ZN2mat}) describes two-dimensional discretized gravity with a single type
of matter state at each of the vertices. It does not admit the
usual $\IZ_2$ Ising symmetry of a classical spin lattice that characterizes the
unitary minimal models of rational conformal field theory \cite{fgz}. Thus the
matter states in the worldsheet interpretation of the fermionic matrix model
are more complicated degrees of freedom than just simple Ising spins or other
conventional types of conformal matter. The logarithmic
nature of the effective two-matrix potential in (\ref{ZN2mat}) is just another
reflection of the connection with Penner matrix models. The original Penner
model \cite{penner} ($g_k=0$ for $k>1$, $g_1=-1$ and $\alpha=1$ in (\ref{ZNH}))
we used to calculate the virtual Euler characteristics of the moduli spaces of
compact Riemann surfaces. For $\alpha>0$ the generalized models are intimately
related to the coupling of gravity to matter fields of central charge $c=1$
\cite{acm}. Note that the Penner interaction in (\ref{ZN2mat}) is well-defined.
We may refer to the
fermionic matrix model as a model of two-dimensional quantum gravity
interacting with ``Penner matter''. As we will see later on, it is this
worldsheet model which leads to a well-defined nonperturbative theory of
quantum gravity in two-dimensions which we will call ``fermionic gravity''.

The partition function (\ref{ZN2mat}) can be written in terms of a double
eigenvalue distribution by diagonalizing the Hermitian matrices $\phi=U\,{\rm
diag}(\phi_1,\dots,\phi_N)\,U^\dagger$ and $\lambda=V\,{\rm
diag}(\lambda_1,\dots,\lambda_N)\,V^\dagger$ by unitary transformations, where
$\phi_i,\lambda_i\in\IR$ are the eigenvalues of $\phi,\lambda$. Computing the
Jacobian for the change of integration measure and using the Itzykson-Zuber
formula \cite{iz} to integrate out the unitary degrees of freedom, we arrive at
the eigenvalue model
\beq
Z_N=c_N\prod_{i=1}^N\int\!\!\!\!\int\limits_{-\infty}^\infty
d\phi_i~d\lambda_i~\lambda_i^N~\e^{NV(\phi_i)+iN\lambda_i
\phi_i}~\Delta(\phi_1,\dots,\phi_N)\,\Delta(\lambda_1,\dots,\lambda_N)
\eeq
where $c_N=(-iN^2)^{N(N+1)/2}/(2\pi)^N\prod_{n=1}^Nn!$ and
\beq
\Delta(x_1,\dots,x_N)=\det_{i,j}\left[x_i^{j-1}\right]=\prod_{i<j}(x_i-x_j)
\label{vandermonde}\eeq
is the Vandermonde determinant. Using
(\ref{vandermonde}) we can then write the partition function as
\beq
Z_N=c_NN!~\det_{i,j}\left[\Bigl(\phi^{i-1}\,,\,\lambda^{j-1}
\Bigr)\right]
\label{zmu}\eeq
where we have introduced the inner product
\beq
\Bigl(F(\phi)\,,\,G(\lambda)\Bigr)\equiv\int\!\!\!\!\int
\limits_{-\infty}^\infty d\phi~d\lambda~\lambda^N~\e^{NV(\phi)+iN\lambda\phi}
{}~F(\phi)\,G(\lambda)
\label{muinnerprod}\eeq
on the vector space $\IC[\phi]\otimes\IC[\lambda]$ of complex coefficient
polynomials in $(\phi,\lambda)$. A remarkable feature of the fermionic matrix
model is the extent to which its inner product (\ref{muinnerprod}) can be
computed. For arbitrary polynomial potential, integrating over $\lambda$ first
in (\ref{muinnerprod}) leads to the result
\beq
\Bigl(F(\phi)\,,\,\lambda^k \Bigr)= \frac{2 \pi}{N}\,
\left( \frac{i}{N}\,\frac{\partial}{\partial\phi}\right)^{N+k}
\left[ F(\phi)~\e^{N V(\phi) } \right] \biggm|_{\phi=0}
\label{master}\eeq
which is valid for any analytic function $F(\phi)$ and any integer $k\geq0$.
Using (\ref{master}), the representation (\ref{zmu}) of the partition function
as the determinant of the moment matrix of the inner product
(\ref{muinnerprod}) thereby yields
\beq
Z_N=(-1)^{[N]_2}\,\det_{i,j}\left[\left(\frac\partial{\partial\phi}+NV'(\phi)
\right)^{N+j-i}\cdot1\biggm|_{\phi=0}\right]
\label{ZNpertexpl}\eeq
where, for any integer $K\geq1$, $[N]_K$ denotes the largest integer less than
or equal to $N/K$. Here and in the following a prime denotes differentiation.
The expression (\ref{ZNpertexpl}) is the desired explicit expansion of the
determinant operator in (\ref{ZNdet}).

\subsection{Combinatorics of Fermionic Ribbon Graphs}

In contrast to the determinant form (\ref{ZNdet}), the representation
(\ref{ZNpertexpl}) leads to an explicit expression for the perturbation series
of the fermionic one-matrix model. For a polynomial potential (\ref{pot}) of
order $K$, the determinant in (\ref{ZNpertexpl}) may be evaluated explicitly by
expanding the exponential function in its Taylor series and applying the
multinomial theorem to each term in the expansion to get
\bea
\frac{\partial^L}{\partial\phi^L}~\e^{NV(\phi)}
\biggm|_{\phi=0}&=&(Ng_1)^L\,\sum_{k_2=0}^{[L]_2}\cdots\sum_{k_K=0}^{[L]_K}
\frac{L!}{(L-2k_2-\cdots-Kk_K)!\,k_2!\cdots k_K!}\nn\\&
&\times\,\prod_{\ell=2}^K\left(\frac{g_\ell}{\ell
N^{\ell-1}\,(g_1)^\ell}\right)^{k_\ell}
\label{Kpotid}\eea
for any integer $L\geq0$. Normalizing by the Gaussian model for which
$V(\phi)=g_1\phi$ and $Z_N^{\rm Gauss}=(-1)^{[N]_2}\,(Ng_1)^{N^2}$, the
partition function
(\ref{ZNpertexpl}) may be written as
\beq
\frac{Z_N}{Z_N^{\rm Gauss}}=\det_{i,j}\left[\sum_{k_2=0}^{[N+j-i]_2}\cdots
\sum_{k_K=0}^{[N+j-i]_K}\frac{(N+j-i)!}{\left(N+j-i-\sum_\ell\ell k_\ell
\right)!\,k_2!\cdots k_K!}\prod_{\ell=2}^K\left(\frac{g_\ell}{\ell N^{\ell-1}\,
(g_1)^\ell}\right)^{k_\ell}\right]
\label{FNpertexpl}\eeq
We can extract the sums in (\ref{FNpertexpl}) out of the determinant by using
the multilinearity of the determinant as a function of its $N$ rows to get
\bea
\frac{Z_N}{Z_N^{\rm Gauss}}&=&\sum_{k_2^{(1)},\dots,
k_2^{(N)}\geq0}\cdots\sum_{k_K^{(1)},\dots,
k_K^{(N)}\geq0}\left[\prod_{m=2}^K\,\prod_{n=1}^N\frac1{k_m^{(n)}!}\,
\left(\frac{g_m}{mN^{m-1}\,
(g_1)^m}\right)^{k_m^{(n)}}\right]\nn\\&
&\times\,\det_{i,j}\left[\frac{(N+j-i)!}{\left(N+j-i-\sum_\ell\ell
k_\ell^{(i)}\right)!}
\,\prod_{\ell=2}^K\Theta\left([N+j-i]_\ell-k_\ell^{(i)}\right)\right]
\label{FNpertexplfin}\eea
where $\Theta$ denotes the step function with the convention $\Theta(0)=1$.

The expression (\ref{FNpertexplfin}) leads to a relatively straightforward
expansion of the partition function in powers of the coupling constants
$g_\ell$, $\ell=2,\dots,K$. It represents the formal solution to the problem of
counting the (connected and disconnected) Feynman-'t~Hooft diagrams of the
adjoint fermion one-matrix model. It is a closed formula, in terms of a sum
over a large set of integers, for the generating function of ribbon graphs of
the fermionic matrix model. A fermionic ribbon graph has much more structure to
it than a conventional fat-graph, and so we shall now formulate the rules for
generating them from the matrix integral. The fermionic propagator is
$\langle\bar\psi_{ij}\psi_{kl}\rangle_{\rm
Gauss}=\delta_{il}\,\delta_{jk}/Ng_1$ which we represent in the usual way as a
double line, along with an orientation defined by an arrow pointing towards the
$\bar\psi$ matrix. Vertices are likewise given an orientation by assigning an
outgoing arrow for a $\bar\psi$ line and an incoming arrow for a $\psi$ line.
Fat-graphs are now drawn with the rule that only $\bar\psi$ and $\psi$ lines
can contract. Each such graph is dual to some discretization of a Riemann
surface in the standard way. The surface may be two-coloured by shading the
triangles in a polygon if and only if their outer edge is dual to a propagator
with an incoming arrow. In this way black edges are always associated with
$\bar\psi$ matrices and white ones with $\psi$. Since the fermionic propagator
only connects $\bar\psi$'s with $\psi$'s, it follows that every such
discretization can be two-coloured. Thus the fat-graphs generated by the
fermionic matrix integral can be obtained by drawing all discretizations
associated with fermionic $k$-point vertices in terms of the corresponding
Hermitian $2k$-point vertices, and keeping only those graphs which are
two-colourable. The even parity of the Hermitian potential $V(\phi^2)$ is
required because only discretizations with even-sided polygons can be
two-coloured.

However, the Grassmann nature of the generating matrices yields important
changes to the rules for counting such triangulations. For a given vertex in
the Wick expansion of the matrix integral, we fix a chosen $\bar\psi$ line.
Lines are joined into propagators with a left-handed orientation. Every time a
$\psi$ line is joined into a $\bar\psi$ line (rather than the other way
around), we ``twist'' the ribbon and thereby reverse its orientation. To each
such twisted line, we associate a factor of $-1$. This standard sign factor for
fermionic fields yields significant reductions in the overall number of ribbon
graphs which are actually generated by the fermionic matrix model, because of
the cancellations which occur, for example, between fermionic diagrams that are
topologically the same but which are twisted relative to one another. Thus
topologically equivalent diagrams do not necessarily add up, but may have the
effect of cancelling each other. As a simple example, consider the quadratic
potential, $K=2$. There are two graphs which contain only a single four-point
vertex, but they contribute with equal magnitude and opposite sign to the
generating function. This vanishing contribution is the coefficient of the
$g_2/2N(g_1)^2$ term in the perturbative expansion of the partition function.
Symbolically, it is given by the Wick expansion of the Gaussian correlator
\beq
\unitlength=1.00mm
\linethickness{0.4pt}
\thicklines
\begin{picture}(100.00,14.00)
\put(0.00,7.00){\makebox(0,0)[l]{$\left\langle\tr(\ps2)^2\right\rangle
_{\rm Gauss}~=$}}
\put(40.00,3.50){\circle{7.00}}
\put(39.00,0.00){\makebox(0,0)[l]{$\scriptstyle{<}$}}
\put(39.00,14.00){\makebox(0,0)[l]{$\scriptstyle{>}$}}
\put(40.00,10.50){\circle{7.00}}
\put(43.00,7.00){\makebox(0,0)[l]{$~~~-~~~$}}
\thicklines
\put(58.0,7.00){\circle{7.00}}
\put(57.00,10.50){\makebox(0,0)[l]{$\scriptstyle{<}$}}
\put(64.00,10.50){\makebox(0,0)[l]{$\scriptstyle{>}$}}
\put(65.00,7.00){\circle{7.00}}
\put(68.00,7.00){\makebox(0,0)[l]{$~~~=~~0$}}
\end{picture}
\label{quadcombs}\eeq
Note the twist in the second propagator of the second graph in
(\ref{quadcombs}), which induces the relative minus sign. Notice also that the
usual toroidal four-point graph does not appear in (\ref{quadcombs}), because
it cannot be two-coloured. This vanishing contribution is also easily found
from the analytical formula (\ref{FNpertexplfin}). The leading term is 1 and it
comes from the configuration whereby all $k$'s are 0. The terms of order $g_2$
come from the configuration whereby only one $k$ is non-vanishing and equal to
unity. In the determinant in (\ref{FNpertexplfin}), the only $\Theta$'s which
are not equal to unity are those which appear in row $i$ with $k^{(i)}=1$. The
determinant therefore vanishes, reproducing the graphical result
(\ref{quadcombs}). The remaining fat-graph combinatorics are now readily
determined from the graphs of the Hermitian one-matrix model with even
potential $V(\phi^2)$ by keeping only those which are two-colourable and
incorporating the appropriate twists. The determinant formula
(\ref{FNpertexplfin}) is readily seen to reproduce the correct combinatorics,
with appropriate minus signs coming from the determinant\footnote{See \cite{pzj} for 
related considerations.}.

These arguments readily generalize to arbitrary polynomial potentials.
Generally, the total number of $2\ell$-valent vertices in a given ribbon graph
$\Gamma$ as determined by the formula (\ref{FNpertexplfin}) is the power of
$g_\ell$ which is given by
\beq
n_{2\ell}(\Gamma)=\sum_{n=1}^Nk_\ell^{(n)}
\label{numvertices}\eeq
and, because of Fermi statistics, this number is bounded as
\beq
0\leq n_{2\ell}(\Gamma)\leq\sum_{n=1}^N[2N-n]_\ell
\label{numbd}\eeq
It follows that there are only a finite number of fermionic fat-graphs
generated. The twisting of ribbons and also the finiteness of the perturbation
series may be attributed to the inclusion of a logarithmic interaction
$-2\log\phi$ in the corresponding Hermitian one-matrix model with potential
$V(\phi^2)$. This role of the Penner potential will be demonstrated explicitly
in the next subsection.

The determinant expansion (\ref{FNpertexplfin}) is therefore an analytic
expression of the combinatorial formula
\beq
\frac{Z_N}{Z_N^{\rm
Gauss}}=\sum_\Gamma\frac{(-1)^{t(\Gamma)}\,(Ng_1)^{\chi(\Gamma)}}{|{\rm
Aut}(\Gamma)|}\,\prod_{\ell=2}^K\left(\frac{g_\ell}{g_1}
\right)^{n_{2\ell}(\Gamma)}
\label{FNcomb}\eeq
where the sum runs through all (connected and disconnected) two-colourable
fat-graphs $\Gamma$ with $n_{2\ell}(\Gamma)$ vertices of order $2\ell$ bounded
as in (\ref{numbd}), and $\chi(\Gamma)$ is the Euler characteristic of
$\Gamma$. The alternating sign factor in (\ref{FNcomb}), with
$t(\Gamma)$ the total number of twisted ribbons of
$\Gamma$, is due to Fermi statistics, while ${\rm Aut}(\Gamma)$ is the
automorphism group of the unmarked graph $\Gamma$. The sum over fat-graphs in
(\ref{FNcomb}) is finite, since the maximum number of vertices that a given
diagram can have is
\beq
v_{\rm max}=\sum_{\ell=2}^K\,\sum_{n=1}^N[2N-n]_\ell
\label{vmax}\eeq
Each such ribbon graph is dual to a tessellation $\Gamma^*$ of a Riemann
surface of Euler characteristic $\chi(\Gamma)$ by $n_{2\ell}(\Gamma)$
$2\ell$-valent tiles for $\ell=2,\dots,K$. Because of the twist factors, the
overall combinatorial numbers generated by the fermionic matrix integral will
be drastically reduced. Moreover, depending on the precise details of the
potential $V$, the perturbative or topological expansions of the matrix model
may be alternating series. The above considerations show that there is clearly
no simple mapping between the Hermitian and fermionic matrix models. These
features will have remarkable implications later on for the topological
expansion of the matrix integral (\ref{1mat}).

\subsection{Unitary Representation}

We will now derive an alternative ``dual'' representation of the perturbative
expansion of the fermionic partition function. For this, we note that the
derivatives in (\ref{master}) can be evaluated in
terms of a contour integration about the origin of the complex plane as
\beq
\Bigl(F(\phi)\,,\,\lambda^k\Bigr)=
- \left( \frac{i}{N} \right)^{N+k+1}\,(N+k)!\,\oint\limits_{z=0}dz~
\frac{F(z)~\e^{N V(z)} }{z^{N + k+1}}
\label{mastercont}\eeq
The partition function (\ref{ZNpertexpl}) may then be written as
\beq
Z_N=c_NN!\,\prod_{k=1}^N \left( - \left( \frac{i}{N} \right)^{N+k}\,(N+ k -1)!
\right) ~\det A
\label{ZNcontint}\eeq
where we have defined the $N\times N$ matrix
\beq
A_{ij}=\oint\limits_{z=0} dz ~\frac{ \e^{N V(z)} }{z^{N + j - i +1}}
\label{Aijdef}\eeq
The crucial observation now is that the matrix elements
(\ref{Aijdef}) depend only on the differences $j-i$
of row and column positions, i.e. $A_{ij}=A_{j-i}$. Matrices with such a
symmetry are known as Toeplitz matrices and their determinants can be evaluated
in terms of averages over the unitary group \cite{itz}. The Toeplitz
determinant
$\det_{i,j}[A_{j-i}]$ appearing in (\ref{ZNcontint}) is associated with the
Laurent series of a function $\cal A$ through the definition
\beq
{\cal A}(\omega) = \sum _{n=-\infty}^{\infty}A_n\,\omega^n
\label{fomega}\eeq
The matrix elements $A_n$ are then interpreted as the Fourier coefficients of
$\cal A$ on the unit circle $|\omega|=1$,
\beq
A_n=\oint\frac{d\omega}{2\pi i\omega}~\omega^{-n}\,{\cal A}(\omega)
\label{anfourier}\eeq
Using the fact that a determinant is a linear function of each of its rows, it
follows that
\beq
\det A=\prod_{k=1}^N\oint\frac{d\omega_k}{2\pi i\omega_k}~{\cal
A}(\omega_k)~\omega_k^k~\overline{\Delta(\omega_1,\dots,\omega_N)}
\label{detavan}\eeq
Replacing the right-hand side of (\ref{detavan}) by its average over all
permutations of the integration variables $\omega_k$ yields
\beq
\det A=\frac1{N!}\,\prod_{k=1}^N\oint\frac{d\omega_k}{2\pi
i\omega_k}~{\cal A}(\omega_k)~\Bigl|\Delta(\omega_1,\dots,\omega_N)\Bigr|^2
\label{detavanperm}\eeq
which shows that the Toeplitz determinant is given by the unitary matrix
integral
\beq
\det A=\frac{\prod_{n=1}^{N-1}n!}{(2\pi)^{N(N-1)/2}}\,
\int\limits_{U(N)}[dU]~\det{\cal A}(U)
\label{toeplitzunitary}\eeq

Upon substituting (\ref{Aijdef}) into (\ref{fomega}), we can interchange the
sum and contour integration provided
we take into account that the latter vanishes for $N+n<0$. We then have
\beq
{\cal A}(\omega) = \oint\limits_{z=0} dz~ \frac{ \e^{N V(z)} }{\omega^N ( z-
\omega)}
= 2 \pi i~\e^{N\bigl( V(\omega) - \log{\omega}\bigr)}
\label{fexplZN}\eeq
{}From (\ref{ZNcontint}), (\ref{toeplitzunitary}) and (\ref{fexplZN}) it
follows that the partition function of the adjoint fermion one-matrix model
(\ref{1mat}) is completely equivalent, for any dimension $N$, to the $N\times
N$ unitary matrix model
\beq
Z_N = k_N \int\limits_{U(N)}[dU]~\e^{N\,\tr\bigl( V(U) - \log{U}\bigr)}
\label{unitaryZN}\eeq
where $k_N=N^{N(N+1)/2}/(2\pi)^{N(N-1)/2}\prod_{n=1}^{N}(N+n-1)!$. We see that
the fermionic one-matrix model for any $N$ is completely equivalent
to both a Hermitian two-matrix model, and a unitary one-matrix model, both of
which involve Penner-type interactions. In particular, the unitary
representation clearly demonstrates that the equivalence with the Hermitian
Penner one-matrix model is not true at finite $N$, but rather only at
$N=\infty$. Note that the matrix integral (\ref{unitaryZN}) is perfectly
well-defined at finite $N$, but it involves an integrand which is not a
real-valued potential.

With the partition function given as an integral over the unitary group, we can
clarify the geometric role of the Penner-type potential that characterizes
fermionic matrix models, and also give an alternative method for a perturbative
expansion in terms of the coupling constants in the potential $V$. For this, we
integrate over the $U(1)$ factor of the unitary group $U(N)=U(1)\times
SU(N)/\IZ_N$ and leave an $SU(N)$ integral,
\beq
Z_N = k_N \int\limits_{U(N)}d\theta~[dU_0]~\e^{N\bigl(\tr\,V(\e^{i
\theta}\,U_0) -i N
\theta\bigr)}=\frac{k_N}N\,\int\limits_0^{2\pi}d\theta~\e^{-i N^2
\theta}\,\int\limits_{SU(N)}[dU_0]~ \e^{N\,\tr\,V(\e^{i \theta}\,U_0) }
\label{ZNU1SUN}\eeq
The role of the $U(1)$ integration over $\theta$ is to restrict the terms in
the $SU(N)$ integration to those involving only $N^2$ factors of $U_0$.
Equivalently, we may expand the invariant function which is the integrand of
(\ref{ZNU1SUN}) in $U(N)$ characters $\chi_{\vec n}$ as
\beq
\e^{N\,\tr\,V(U)}=\sum_{\vec n}c_{\vec n}~\chi_{\vec
n}(U)~~~~~~,~~~~~~U=\e^{i\theta}\,U_0
\label{charexp}\eeq
where $c_{\vec n}\in\IC$ are functions of the coupling constants of the
potential $V$, and the sum goes over all irreducible polynomial representations
of $U(N)$ which may be parameterized by their highest weight components
$n_1\geq\dots\geq n_N\geq0$ associated with the lengths of the lines in the
corresponding Young tableaux. Since
\beq
\chi_{\vec n}(U)=\e^{i\theta C_1(\vec n)}\,\chi_{\vec n}(U_0)
\label{chiU}\eeq
where
\beq
C_1(\vec n)=\sum_{k=1}^Nn_k
\label{C1vecn}\eeq
is the linear Casimir of the $U(N)$ representation $\vec n$ (the total number
of boxes in the Young tableau), it follows that the constraint on the $SU(N)$
integration in (\ref{ZNU1SUN}) is a restriction to terms with linear $U(N)$
Casimir $C_1$ equal to $N^2$. We may therefore write
\beq
Z_N
=\frac{k_N}N~\left[~\int\limits_{SU(N)}[dU_0]~\e^{N\,\tr\,V(U_0)}\,
\right]_{C_1=N^2}
\label{ZNrestr}\eeq
where the bracket means to restrict the character expansion
of $\e^{N\,\tr\,V(U_0)}$ to those $\vec n$ with $C_1(\vec n)=N^2$. The role of
the logarithmic interaction in (\ref{unitaryZN}) may therefore be characterized
as restricting the matrix integral to completely filled Young diagrams.

The $SU(N)$ integral in (\ref{ZNrestr}) can be evaluated using standard
techniques from lattice gauge theory \cite{creutz}. For this, we introduce the
generating function
\beq
W(J)=\int\limits_{SU(N)}[dU_0]~\e^{\tr\,J\,U_0}
\label{genfndef}\eeq
with the property
\beq
\int\limits_{SU(N)}[dU_0]~f(U_0)=f\left(\mbox{$\frac{\partial}{\partial J}$}
\right)W(J)\biggm|_{J=0}
\eeq
where $J$ is an arbitrary $N\times N$ matrix and $f$ an arbitrary function on
$SU(N)$. Using the invariance of the Haar measure $[dU_0]$ under $SU(N)$
rotations, we have $W(J)=W(UJV)$, $U,V\in SU(N)$, from which it can be shown
that the generating function may be expanded in powers of determinants of $J$
as \cite{creutz}
\beq
W(J)=\frac{(2\pi)^{\frac{N(N+1)}2-1}}N\,\sum_{k=0}^\infty\frac{(\det
J)^k}{\prod_{n=1}^N(n+k-1)!}
\label{WJfinal}\eeq
Applying this result to the partition function (\ref{ZNrestr}), the restriction
$C_1(\vec n)=N^2$ implies that only the $k=N$ term of the expansion
(\ref{WJfinal}) contributes, giving
\beq
Z_N =\e^{   N\,\tr\,V \bigl( \frac{\partial}{\partial J} \bigr)  }~
{\det}^N{J}\biggm|_{J=0}
\label{ZNdetunitary}\eeq
Comparing (\ref{ZNdet}) and (\ref{ZNdetunitary}), we see that the finite
$N$ expressions for the fermionic partition function in the Hermitian
two-matrix and unitary one-matrix representations are dual to each other. We
can make this duality precise by writing (\ref{ZNdetunitary}) as
\bea
Z_N &=&\frac1{(2 \pi)^{N^2}}\,\int\!\!\!\!\!\int\limits_{gl(N)}dJ~d\phi~
\e^{i\,\tr\,\phi J}{}~\e^{   N\,\tr\,V \bigl( \frac{\partial}{\partial J}
\bigr)  }~{\det}^N{J}\nonumber\\& = &-\frac1{(2 \pi)^{N^2}}
\,\int\!\!\!\!\!\int\limits_{gl(N)}dJ~d\phi~
\e^{   N\,\tr\,V ( i \phi)  }~{\det}^N\left(\mbox{$-i\,
\frac{\partial}{\partial\phi}$}\right) ~\e^{i\,\tr\,\phi J}\nonumber\\
& = &\int\limits_{gl(N)}d\phi~\delta(\phi) ~
{\det}^N\left(\mbox{$-i\,\frac{\partial}{\partial\phi}$}\right)~\e^{N\,\tr\,V
( i \phi)  }~=~
{\det}^N\left(\mbox{$\frac{\partial}{\partial\phi}$}\right)~\e^{N\,\tr\,V
(\phi)  } \biggm|_{\phi=0}
\label{ZNdetdual}\eea
Some further aspects of the unitary representation of the fermionic matrix
model are discussed in appendix A. In section 5 we will show explicitly that
the equations of motion of the unitary and fermionic one-matrix models are
identical.

\section{Orthogonal Polynomial Solution}
\setcounter{equation}{0}

In this section we shall demonstrate how to generalize the orthogonal
polynomial technique \cite{biz} to the adjoint fermion one-matrix model
(\ref{1mat}), with the goal of solving the matrix model in the large $N$ limit.
Although the fermionic matrix model is not naturally an eigenvalue model, since
it is not possible to diagonalize a Grassmann valued matrix, we can exploit its
equivalence with the unitary one-matrix model (\ref{unitaryZN}). We will
thereby {\it define} the orthogonal polynomials of the fermionic matrix model
(\ref{1mat}) to be the orthogonal polynomials associated with the corresponding
unitary eigenvalue model. This definition will lead, as we shall see, to a
well-defined real-valued solution for the fermionic matrix model.

\subsection{General Properties}

Let us consider the fermionic partition function in the representation
(\ref{ZNcontint},\ref{Aijdef}), which involves the measure
\beq
d\mu(z)=\frac{\e^{NV(z)}}{z^{N+1}}~dz
\label{meas}\eeq
on the unit circle $z\bar z=1$. The partition function (\ref{ZNcontint}) may
then be written as the determinant of the moment matrix of the measure
(\ref{meas}),
\beq
Z_N=d_N~\det_{i,j}\left[\Bigl\langle z^{i-1}\Bigm|\bar
z^{j-1}\Bigr\rangle\right]
\label{ZNunitarydet}\eeq
where $d_N$ is an irrelevant numerical constant and we have defined the inner
product
\beq
\Bigl\langle F(z)\Bigm|G(\bar z)\Bigr\rangle=\oint\limits_{z=0}
d\mu(z)~F(z)\,G(\bar z)
\label{unitaryinnprod}\eeq
on the space $\IC[z]\otimes\IC[\bar z]$ of Laurent polynomials on the unit
circle. The construction of orthogonal polynomials to evaluate inner products
on the circle such as (\ref{unitaryinnprod}) with real, positive definite
measure $d\mu$ was discussed long ago in the mathematics literature
\cite{szego} and subsequently in the physics literature \cite{genumm}. Some
aspects of generic unitary matrix models are also dealt with in
\cite{genunitary,adler}. In the present case, the measure (\ref{meas}) is
complex-valued, and we must deal
accordingly with defining the appropriate system of orthogonal polynomials.

These are the monic polynomials $(\Phi_n(z),\Lambda_m(z))$ of order $(n,m)$
which form a complete set in the space $\IC[z]\otimes\IC[\bar z]$ and which
are bi-orthogonal in the measure (\ref{meas}),
\beq
\Bigl\langle\Phi_n(z)\Bigm|\Lambda_m(z)\Bigr\rangle=h_n\delta_{nm}
\label{ortho}\eeq
where $h_n$ are some constants. They are normalized as
\beq
\Phi_n(z)=z^n-\sum_{k=0}^{n-1}p_{n,k}\,z^k~~~~~~,~~~~~~
\Lambda_m(z)=z^{-m}-\sum_{k=0}^{m-1}l_{m,k}\,z^{-k}
\label{norm}\eeq
where the coefficients $p_{n,k}$ and $l_{m,k}$ can be formally obtained from
the usual Gram-Schmidt orthogonalization procedure applied to the monomials
$(z^n,z^{-m})$ and the inner product (\ref{unitaryinnprod}). In fact, by
iterating (\ref{norm}) we find that the monomials $(z^n,z^{-m})$ can be
expanded as
\bea
z^n&=&\Phi_n(z)+\sum_{j=0}^n\left(\sum_{k_1=0}^{n-1}\,\sum_{k_2=0}^{k_1-1}
\cdots\sum_{k_j=0}^{k_{j-1}-1}p_{n,k_1}p_{k_1,k_2}\cdots p_{k_{j-1},k_j}
\,\Phi_{k_j}(z)\right)\nonumber\\z^{-m}&=&\Lambda_m(z)+\sum_{j=0}^m\left(
\sum_{k_1=0}^{m-1}\,\sum_{k_2=0}^{k_1-1}
\cdots\sum_{k_j=0}^{k_{j-1}-1}l_{m,k_1}l_{k_1,k_2}\cdots l_{k_{j-1},k_j}
\,\Lambda_{k_j}(z)\right)
\label{monicexp}\eea
It follows that any polynomial in the space $\IC[z]\otimes\IC[\bar z]$ of
degree $(n,m)$ can be expressed as a linear combination of
$(\Phi_k(z),\Lambda_\ell(z))$ with $k\leq n$ and $\ell\leq m$. From
(\ref{norm}) it follows they define a change of basis
$(z^{k-1},z^{-\ell+1})\mapsto(\Phi_{k-1}(z),\Lambda_{\ell-1}(z))$ of the vector
space $\IC[z]\otimes\IC[\bar z]$ that leaves the partition function
(\ref{ZNunitarydet}) invariant and at the same time diagonalizes the inner
product (\ref{unitaryinnprod}). In particular, from (\ref{ortho}) we have
\beq
Z_N=d_N~\det_{i,j}\left
[\Bigl\langle\Phi_{i-1}(z)\Bigm|\Lambda_{j-1}(z)
\Bigr\rangle\right]=d_N~h_0^N\,\prod_{n=1}^{N-1}R_n^{N-n}
\label{rec}\eeq
where we have introduced the recursion coefficients
\beq
R_n=\frac{h_n}{h_{n-1}}
\label{reccoeffs}\eeq
and $h_0=\oint_{z=0}d\mu(z)$ is the normalization of the measure (\ref{meas}).

The reason why two independent sets of functions corresponding to clockwise and
anti-clockwise rotations on the circle are required here is the complexity of
the measure (\ref{meas}). In the classical case whereby the measure $d\mu$ is
real and positive \cite{szego}, the polynomials $\Phi_n(z)$ and $\Lambda_n(z)$
are related to each other by complex conjugation. The present
system of polynomials may be thought of as a deformation of those for positive
definite measures on the unit circle. This deformation is essentially encoded
in the coefficients $p_{n+1,0}$ and $l_{n+1,0}$.
In order for the Gram-Schmidt process to
work for the complex measure (\ref{meas}), one needs polynomials in both $z$
and $1/z$ and {\it a priori} these polynomials are unrelated. 
Alternatively, we
can think of this doubling as the usual doubling of degrees of freedom due to
the fermionic nature of the original matrix model. As the orthogonal
polynomials define a complete bi-orthogonal system of functions, they satisfy
the completeness relation
\beq
\sum_{n=0}^\infty\frac{\Phi_n(z)\,\Lambda_n(z)}{h_n}=
\delta^{(\mu)}(z)=z^{N+1}~\e^{-NV(z)}\,\delta(z)
\label{complrel}\eeq
on the vector space $\IC[z]\otimes\IC[\bar z]$, where the Dirac
delta-function corresponding to the measure $d\mu$ is defined so that
\beq
\oint\limits_{z=0} d\mu(z)~F(z)~\delta^{(\mu)}(z-z')=F(z')
\label{deltamyu}\eeq
for any function $F(z)$ on the unit circle. The
relation (\ref{complrel}) is understood as an analytical continuation (to the
distribution space that is the completion of $\IC[z]\otimes\IC[\bar z]$)
which can be thought of as a representation of the potential $V(\ps2)$ in terms
of the orthogonal polynomials.

The problem of evaluating the partition function of the fermionic matrix model
therefore boils down to evaluating the coefficients $R_n$ appearing in
(\ref{rec}). For this, we will first derive a few properties of the system of
orthogonal polynomials introduced above. Completeness implies the relations
\beq
z\,\Phi_n(z)=\Phi_{n+1}(z)+\sum_{k=0}^n{\cal P}_k^{(n)}\,\Phi_k(z)
{}~~~~~~,~~~~~~\frac1z\,\Lambda_m(z)=\Lambda_{m+1}(z)+\sum_{k=0}^m
{\cal L}_k^{(m)}\,\Lambda_k(z)
\label{rels}\eeq
The coefficients ${\cal P}_k^{(n)}$ and ${\cal L}_k^{(m)}$
completely determine the form of the orthogonal polynomials and hence
the partition function. To see this, we use (\ref{norm}) and (\ref{rels}) to
write
\bea
\Phi_n(z)&=&z\,\Phi_{n-1}(z)-\sum_{k=0}^{n-1}{\cal
P}_k^{(n-1)}\,\Phi_k(z)\nn\\&=&z^n-{\cal
P}_0^{(n)}-\sum_{k=1}^{n-1}\left(p_{n-1,k-1}+{\cal
P}_k^{(n-1)}\right)z^k-\sum_{k=0}^{n-2}{\cal
P}_k^{(n-1)}\,\sum_{j=0}^{k-1}p_{k,j}\,z^j
\label{Phincompare}\eea
Comparing the various polynomial coefficients in (\ref{Phincompare}) with those
of the definition (\ref{norm}), we arrive at the iterative relations
\bea
p_{n,n-1}&=&p_{n-1,n-2}+{\cal P}_{n-1}^{(n-1)}\label{pnn-1it}\\p_{n,k}&=&{\cal
P}_k^{(n-1)}+p_{n-1,k-1}+\sum_{j=k+1}^{n-1}p_{j,k}\,{\cal
P}_j^{(n-1)}~~~~~~,~~~~~~1\leq k\leq n-2\label{pnkit}\\p_{n,0}&=&{\cal
P}_0^{(n)}+\sum_{j=1}^{n-1}p_{j,0}\,{\cal P}_j^{(n-1)}
\label{pn0it}\eea
The relations (\ref{pnn-1it}) and (\ref{pn0it}) can be iterated
straightforwardly and by induction we have
\bea
p_{n,n-1}&=&\sum_{j=0}^{n-1}{\cal P}_j^{(j)}\label{pnn-1sol}\\p_{n,0}&=&{\cal
P}_0^{(n)}+\sum_{k=1}^{n-3}\left(\sum_{j_1=1}^{n-1}\,\sum_{j_2=1}^{j_1-1}
\cdots\sum_{j_k=1}^{j_{k-1}-1}{\cal P}_{j_1}^{(n-1)}{\cal P}_{j_2}^{(j_1-1)}
\cdots{\cal P}_{j_k}^{(j_{k-1}-1)}{\cal P}_0^{(j_k)}\right)\nn\\
& &+\,{\cal P}_0^{(1)}{\cal P}_1^{(1)}\sum_{j_1=1}^{n-1}\,\sum_{j_2=1}^{j_1-1}
\cdots\sum_{j_{n-2}=1}^{j_{n-3}-1}{\cal P}_{j_1}^{(n-1)}
{\cal P}_{j_2}^{(j_1-1)}\cdots{\cal P}_{j_{n-2}}^{(j_{n-3}-1)}
\label{pn0sol}\eea
which can be substituted into (\ref{pnkit}) to iteratively determine the
remaining $p_{n,k}$. Analogous relations exist between the coefficients
$l_{m,k}$ and ${\cal L}_k^{(m)}$.

We will also need higher degree versions of the completeness relations
(\ref{rels}). For this, we define constants $P_{n,k}^{[\ell]}$ by
\beq
z^\ell\,\Phi_n(z)=\sum_{k=0}^{n+\ell}P_{n,k}^{[\ell]}~\Phi_k(z)
{}~~~~~~{\rm with}~~P^{[\ell]}_{n,n+\ell}=1
\label{Pdef}\eeq
and use (\ref{rels}) to write
\beq
z^{\ell+1}\,\Phi_n(z)=\sum_{k=0}^{n+\ell+1}P_{n,k}^{[\ell+1]}\,\Phi_k(
z)=\sum_{k=0}^{n+\ell}P_{n,k}^{[\ell]}\left(\Phi_{k+1}(z)+\sum_{j=0}
^k{\cal P}_j^{(k)}\,\Phi_j(z)\right)
\label{compliter}\eeq
Using completeness of the orthogonal polynomials we arrive at a recursive
relation for the coefficients $P_{n,k}^{[\ell]}$,
\beq
P_{n,k}^{[\ell+1]}=P_{n,k-1}^{[\ell]}+\sum_{j=k}^{n+\ell}P_{n,j}^{[\ell]}
{}~{\cal P}_k^{(j)}~~~~~~,~~~~~~0\leq k\leq n+\ell+1
\label{Precrel}\eeq
with $P_{n,-1}^{[\ell]}\equiv0$, ${\cal P}_k^{(n)}\equiv0$ for $k>n$, and the
initial
condition $P_{n,k}^{[0]}=\delta_{nk}$.

\subsection{Recursion Relations}

We will now determine the coefficients appearing in the completeness relations
using the above properties of the orthogonal polynomials. This will lead to a
set of recursion equations for the coefficients $R_n$ which thereby completely
determines the solution of adjoint fermion one-matrix model. Consider the
bi-orthogonality relation
\beq
\Bigl\la\Phi_{n+1}(z)\Bigm|\Lambda_m(z)\Bigr\ra=0~~~~~~,~~~~~~ 0\leq m\leq n
\label{orthorelPhi}\eeq
Using the completeness relations (\ref{rels}) we can write (\ref{orthorelPhi})
as
\beq
\Bigl\la z\,\Phi_{n}(z)\Bigm|\Lambda_m(z)\Bigr\ra={\cal P}_m^{(n)} h_m
\label{zPhiLambda}\eeq
The inner product in (\ref{zPhiLambda}) can be evaluated by grouping the factor
of $z$ with the polynomial $\Lambda_m(z)$ with the result
\beq
\Bigl\la z\,\Phi_{n}(z)\Bigm|\Lambda_m(z)\Bigr\ra=\Bigl\la \Phi_{n}(z)\Bigm|
\Lambda_{m-1}(z)+O\left(z^{-m+2}\right)-l_{m,0}\,z\Bigr\ra=-l_{m,0}\,\Bigl\la
z\,\Phi_{n}(z)\Bigm|1\Bigr\ra
\label{zPhirel}\eeq
where we have used the fact that, aside from the term $l_{m,0}\,z$ which comes
from the constant part of the $\Lambda$ polynomials in (\ref{norm}), the
function $z\,\Lambda_m(z)$ is a sum of polynomials of degree $\leq m-1$. The
final inner product in (\ref{zPhirel}) can be evaluated in a straightforward
fashion
using the relation
\beq
\Bigl\la z\,\Phi_{n}(z)\Bigm|\Lambda_{n+1}(z)\Bigr\ra = h_{n+1}
\eeq
which follows from (\ref{rels}). Again, by grouping the factor of $z$ with
$\Lambda_{n+1}(z)$ one finds
\beq
\Bigl\la z\,\Phi_{n}(z)\Bigm|1\Bigr\ra = \frac{ h_{n} - h_{n+1}}{l_{n+1,0}}
\label{zPhiexpl}\eeq
Substituting (\ref{zPhiexpl}) and (\ref{zPhirel}) into (\ref{zPhiLambda}), we
find a concise relation between the constant terms of the $\Phi$ polynomials
and the completeness coefficients ${\cal P}_m^{(n)}$,
\bea
{\cal P}_m^{(n)}&=&\frac{l_{m,0}}{l_{n+1,0}}\,\frac{ h_{n+1} - h_{n}}{h_m}
\nn\\&=&\left\{\new{\begin{array}{l}Q_m\Bigl(R_{n+1}-1\Bigr)
\prod_{k=m+1}^nQ_kR_k~~~~~~
{\rm for}~~0\leq m<n\\Q_n(R_{n+1}-1)~~~~~~{\rm for}~~m=n\end{array}}\right.
\label{calPQR}\eea
where we have defined
\beq
Q_n = \frac{l_{n,0}}{l_{n+1,0}}
\label{Qndef}\eeq
The completeness relation (\ref{rels}) for the $\Phi$ polynomials can therefore
be written as
\beq
z\,\Phi_n(z)=\Phi_{n+1}(z)+\frac{ h_{n+1} - h_{n}}{l_{n+1,0}}\,
\sum_{k=0}^n \frac{l_{k,0}}{h_k}\,\Phi_k(z)
\eeq
Subtracting this relation from itself under the shift of index $n\rightarrow
n-1$ leads to the three-term recursion relation for the $\Phi$ polynomials
\beq
\Phi_{n+1}(z) =  z\,\Phi_n(z)-Q_n\,\frac{ R_{n+1}- 1}{R_n - 1}\,
\Bigl(\Phi_n(z) - z\,R_n\,\Phi_{n-1}(z)\Bigr)
\label{3termPhi}\eeq

Analogous relations for the $\Lambda$ polynomials may also be derived. Equating
the constant terms on both sides of (\ref{3termPhi}) yields
\beq
\frac{p_{n+1,0}}{p_{n,0}}=-Q_n\,\frac{R_{n+1}-1}{R_n-1}
\label{pratio}\eeq
so that
\bea
{\cal L}_m^{(n)}&=&\frac{p_{m,0}}{p_{n+1,0}}\,\frac{h_{n+1}-h_n}{h_m}\nn\\&=&
\left\{\new{\begin{array}{l}(-1)^{n-m}\,\frac{1-R_m}{Q_m}\,
\frac{R_{n+1}-1}{R_{m+1}-1}\,
\prod_{k=m+1}^n\frac{R_k}{Q_k}\,\frac{R_{k}-1}{R_{k+1}-1}~~~~~~
{\rm for}~~0\leq m<n\\\frac{1-R_n}{Q_n}~~~~~~{\rm for}~~m=n\end{array}}\right.
\label{calLexpl}\eea
Substituting (\ref{calLexpl}) into the completeness relation (\ref{rels}) for
the $\Lambda$ polynomials and subtracting the resulting expression from itself
under the index shift $n\to n-1$, we arrive at the three-term recursion
relation
\beq
\Lambda_{n}(z) =\frac1z\,R_n\,\Lambda_{n-1}(z)-Q_n
\left( \Lambda_{n+1}(z) -\frac1z\,\Lambda_{n}(z)\right)
\label{3termLambda}\eeq

The recursion relations (\ref{3termPhi}) and (\ref{3termLambda}) have been
derived without any reference to the particular details of the model, i.e. they
hold independently of the precise form of the potential $V$
of the fermionic matrix model.  Consequently,
these relations are merely `kinematical' in origin.  In order to solve
for the dynamics of the matrix model we need
more information.  From a more pragmatic point of view, in order
to solve for the partition function we
need to find a solution for the coefficients $R_n$, but the recursion
relations also involve the quantities $Q_n$.  More information
is needed to link these two objects.
As we will see, one only needs two `dynamical' relations to obtain
a closed set of recursion relations for any polynomial potential.
The first one is given from (\ref{ortho}) and (\ref{norm}) as
\beq
\Bigl\la \Phi_n(z)\Bigm|\Lambda_{n-1}^\prime (z)\Bigr\ra = (1-n) h_n
\label{dyn1}\eeq
Integrating by parts on the left-hand side of (\ref{dyn1}) gives
\beq
(N+1)\Bigl\la\Phi_n(z)\Bigm|\mbox{$\frac1z$}\,\Lambda_{n-1}(z)\Bigr\ra
-N\Bigl\la V'(z)\,\Phi_n(z)\Bigm|\Lambda_{n-1}(z)\Bigr\ra
-\Bigl\la\Phi_n'(z)\Bigm|\Lambda_{n-1}(z)\Bigr\ra=(1-n)h_n
\label{dyn1intparts}\eeq
Using the completeness relation (\ref{rels}) for the $\Lambda$ polynomials and
the fact that $\Phi_n'(z)=n\Phi_{n-1}(z)+O(z^{n-2})$, we arrive at
\beq
N\Bigl\la V^\prime(z)\,\Phi_n(z)\Bigm|\Lambda_{n-1}(z)\Bigr\ra = (N+n) h_n - n
h_{n-1}
\label{dyn1expl}\eeq
For a polynomial potential (\ref{pot}) of degree $K$, we may use (\ref{Pdef})
to write (\ref{dyn1expl}) as an equation for the recursion coefficients $R_n$,
\beq
R_n=\frac n{n+N}+\frac N{n+N}\,\sum_{k=1}^{K-1}g_{k+1}\,P_{n,n-1}^{[k]}
\label{Rneqn}\eeq

Closely related to the first one, the second `dynamical' relation is given by
the bi-orthogonality relation
\beq
\Bigl\la \Phi_n(z)\Bigm|\Lambda_{n-2}^\prime (z)\Bigr\ra = 0
\label{dyn2}\eeq
Integrating this by parts gives
\beq
N\Bigl\la V^\prime(z)\,\Phi_n(z)\Bigm|\Lambda_{n-2}(z)\Bigr\ra =
-\Bigl\la \Phi_n^\prime(z)\Bigm|\Lambda_{n-2}(z)\Bigr\ra
\label{dyn2intparts}\eeq
The right-hand side of (\ref{dyn2intparts}) can be evaluated using
(\ref{norm}), (\ref{monicexp}) and orthogonality, and keeping only those terms
in the $\Phi$ polynomials which have a non-vanishing overlap with
$\Lambda_{n-2}(z)$ in (\ref{dyn2intparts}). This gives
\bea
\Bigl\la \Phi_n^\prime(z)\Bigm|\Lambda_{n-2}(z)\Bigr\ra
&=&\Bigl\la n\,z^{n-1}-(n-1) p_{n,n-1}\,z^{n-2}
+O\left(z^{n-3}\right)\Bigm|\Lambda_{n-2}(z)\Bigr\ra\nn\\
& = &\Bigl[(n-1) p_{n,n-1}- n p_{n-1,n-2}\Bigr]h_{n-2}
\label{dyn2overlap}\eea
By defining
\beq
P_n =\frac{p_{n,n-1}}N
\label{Pndef}\eeq
we thus find that the second `dynamical' relation reads
\beq
n P_{n-1} -(n-1) P_{n}=\sum_{k=1}^{K-1}g_{k+1}\,P_{n,n-2}^{[k]}
\label{Pneqn}\eeq
The coefficients $P_n$ can be related to the coefficients $Q_n$ and $R_n$
by using (\ref{pnn-1sol}) and (\ref{calPQR}) to get
\beq
N( P_{n+1} - P_n) = Q_n (R_{n+1}-1)
\label{PQRrel}\eeq

To summarize the results of this subsection, we note that the right-hand sides
of (\ref{Rneqn}) and (\ref{Pneqn}) can be evaluated, using
the recursion formula (\ref{Precrel}), in terms of the
completeness coefficients ${\cal P}_m^{(n)}$. Using (\ref{calPQR}),
it follows that, for any polynomial potential, (\ref{Rneqn}), (\ref{Pneqn}) and
(\ref{PQRrel}) form a closed set of recursion
relations involving only the coefficients $R_n$, $Q_n$ and $P_n$. An explicit
system of fermionic orthogonal polynomials is constructed in appendix B.

\subsection{Free Energy}

The final quantity we need for the solution of the fermionic matrix model
is the normalization $h_0$ of the measure
(\ref{meas}). This integral can be evaluated explicitly to give
\beq
h_0=\oint\limits_{z=0}dz~\frac{\e^{NV(z)}}{z^{N+1}}=\frac{2\pi
i}{N!}~\frac{\partial^N}{\partial z^N}~\e^{NV(z)}\biggm|_{z=0}
\label{h0expl}\eeq
We substitute these quantities into (\ref{rec}) and normalize by the Gaussian
partition function $Z_N^{\rm Gauss}=(-1)^{[N]_2}\,
(Ng_1)^{N^2}$ for which $V(z)=g_1z$,
$R_n=n/(n+N)$ and $h_0=2\pi i(Ng_1)^N/N!$. For a polynomial potential, we may
evaluate (\ref{h0expl}) explicitly using (\ref{Kpotid}), and using the
multinomial theorem we thus find that the free energy
$F_N=\frac{1}{N^2}\log\frac{Z_N}{Z_N^{\rm Gauss}}$ is given by
\bea
F_N&=&\sum_{n=1}^\infty(-1)^n\,(n-1)!\,\sum_{0\leq
k_0^{(2)},\dots,k_{[N]_2}^{(2)}\leq n}\cdots\sum_{0\leq
k_0^{(K)},\dots,k_{[N]_K}^{(K)}\leq n}\frac{\delta
\left(\sum_{\ell,m}k_\ell^{(m)}-n\right)}{\prod_{\ell,m}k_\ell^{(m)}!}
\nn\\& &\times\,\prod_{j=2}^K\,\prod_{\ell_j=1}^{[N]_j}\left[\frac{N!}
{(N-\sum_ii\ell_i)!\,\ell_2!\cdots\ell_K!}\,\prod_{m=2}^K\left(
\frac{g_m}{mN^{m-1}(g_1)^m}\right)^{\ell_m}\right]^{k_{\ell_j}^{(j)}}\nn\\&
&+\,\frac{1}{N}\,\sum_{n=1}^{N-1}\left(1-\frac{n}{N}\right)
\log\frac{(n+N)R_n}{n}
\label{FNrec}\eea
where the coefficients $R_n$ are determined from (\ref{Rneqn}).

Generally, in the large-$N$ limit, the variable $x\equiv n/N$ becomes a
continuous parameter in the interval $[0,1]$. We assume that in this limit the
recursion coefficients ${\cal R}_n$ (any of the quantities $R_n$, $Q_n$ or
$P_n$), determined as the solutions of the recursion equations derived in the
previous subsection in the large-$N$
limit, become continuous functions ${\cal R}(x)$ of $x\in[0,1]$. This is
justified by the form of the recursion relations which suggest the replacement, 
$
{\cal R}_n \rightarrow{\cal R}(\mbox{$\frac{n}{N}$}) ={\cal R}(x)
$.
It follows that shifts in the index $n$ are related to sub-leading
contributions in the large $N$ limit,
\beq
{\cal R}_{n+a} \rightarrow{\cal R}(x) +
\frac{a}{N}\,\sum_{k=1}^\infty\frac1{k!}\,\left(\frac
aN\right)^{k-1}\,\frac{\partial^k
{\cal R}(x)}{\partial x^k}
\label{calRreplace}\eeq
and the second, finite sum in (\ref{FNrec}) can be approximated by a simple
one-dimensional integral over $x\in[0,1]$ at $N=\infty$. To deal with the first
term in (\ref{FNrec}), we need to determine the large $N$ limit of the
normalization constant $h_0$. For this, we note that there is nothing
particularly special about the choice of integration contour in (\ref{h0expl}),
as it was merely introduced in (\ref{mastercont}) as a means of evaluating the
derivatives in (\ref{master}). In particular, since its integrand is an
analytic function, we can deform the contour arbitrarily in the complex plane,
and in particular to one out at infinity. In the large $N$ limit, the
one-dimensional integral $h_0$ may then be evaluated using the saddle-point
approximation. In this way we find that the genus 0 free energy is given by
\beq
F_0\equiv\lim_{N\to\infty}F_N=f_0+\int\limits_0^1dx~(1-x)
\log\frac{(1+x)R(x)}{x}
\label{F0}\eeq
where
\beq
f_0=V(\zeta)-\log g_1\zeta
\label{f0}\eeq
with $\zeta$ the solution of the order $K$ algebraic saddle-point equation
\beq
\zeta\,V'(\zeta)=1
\label{zetaeq}\eeq
for the branch which has the perturbative Gaussian limit
$\zeta|_{g_2=\dots=g_K=0}=1/g_1$ and for which the function (\ref{f0}) is
real-valued.

\section{Topological Expansion}
\setcounter{equation}{0}

In this section we will solve for the continuum limit of the discretized random
surface theory represented by the adjoint fermion one-matrix model. We will
consider two limits of the matrix model. The first one is the naive large-$N$
limit which captures the leading critical behaviour and gives a
non-perturbative solution to the problem of counting fermionic ribbon graphs of
spherical topology. The second one is the double-scaling limit in which an
appropriate coupling constant is tuned in the limit $N\to\infty$ in order to
obtain contributions from all orders of the $\frac1N$ expansion of the free
energy and which yields the complete topological expansion of the fermionic
matrix model.

\subsection{Critical Behaviour}

We will start by making some general remarks concerning the large $N$ limit of
the fermionic matrix model. From (\ref{F0}) we see that there are two
contributions to the $N=\infty$ free energy. These two quantities represent
very different phases of the system. The first one is given by (\ref{f0}) which
is determined by the saddle-point equation (\ref{zetaeq}). This equation is
immediately recognized as the critical equation describing the continuum limit
of an abstract branched polymer \cite{2Dbook,BP}. In fact, the explicit
derivative expression for the measure normalization (\ref{h0expl}) coincides
exactly with the partition function of the $N$ dimensional fermionic {\it
vector} model with the same polynomial potential $V$ \cite{ss2}. The latter
quantity is related to the generating function for branched polymer networks by
a simple mapping of its Feynman diagrams and an analytical continuation
$N\mapsto-\frac N2$ of the fermionic vector dimension. The analytic reason for
the appearance of branched polymer behaviour in the fermionic matrix model is
that the normalization of its measure is not of sub-leading order in $N$. From a
graphical point of view, its appearance is clear from the analysis of section
2.2. The entropy factors
associated with the number of Feynman graphs is drastically reduced in the
fermionic case because of the cancellations which occur between twisted
diagrams. There is a class of diagrams with no twists in the fermionic matrix
model, and these have a tree-like growth in the large $N$ limit. The reduced
entropy of the remaining twisted graphs is then comparable to that of the
graphs which produce a polymer-like behaviour. This feature is unique to
fermionic matrices, and it is the property which enables the construction of a
model of branched polymers using supersymmetric matrix models \cite{amzpoly}
via the coupling of the fermionic matrix model to an ordinary, complex matrix
model which has the effect of cancelling the set of twisted diagrams. In fact,
it is precisely the twisting mechanism described in section 2.2 that enables
one to isolate graphs with tree-like growth in the fermionic case. The
untwisted diagrams may then be mapped onto branched polymers similarly to the
case of the cactus diagrams which appear in supersymmetric matrix models
\cite{amzpoly}.

The twisted graphs are associated with the second contribution in (\ref{F0})
and, as we will demonstrate, they lead to the usual surface effects in the
theory. Thus the orthogonal polynomial formalism that we have developed
naturally splits the generating function into a piece corresponding to the
tree-like graphs and a piece corresponding to the twisted diagrams. The
continuum limit of the latter ensemble of graphs describes a surface theory and
describes an effect which is of order $N^2$, while the former ensemble which
has a polymer growth produces an effect of order $N$. One typical feature of
the polymer generating function is that it becomes non-analytic in the large
$N$ limit at the Gaussian point $g_k=0$, $k\geq2$. However, this effect is of
sub-leading order in $N$, so that a large $N$ analysis with arbitrary coupling
constants will miss the branched polymer phase transition. There is a
``barrier'' at the Gaussian point which separates the surface-like continuum
limit from the branched polymer behaviour. From a geometrical point of view,
the fermionic matrix model probes an intermediate lattice phase where the
number of Feynman diagrams contributing to a given process is greater than that
expected of a vector model but less than that of a matrix model. Thus the
matrix integral generates a random surface theory which contains an
``internal'' branched polymer phase.

This decrease in the number of graphs as compared to the usual matrix models
will be responsible for the Borel summability of the all genus expansion of the
free energy of the fermionic matrix model that we will prove in this section.
As discussed above, the essential critical behaviour of the fermionic matrix
model is encoded in the large-$N$ limit of the recursion coefficients $R_n$. If
we introduce the quantities $\tau_n$ defined by
\beq
h_n=\frac{\tau_{n+1}}{\tau_n}
\label{taundef}\eeq
with $\tau_0=d_N$ as in (\ref{ZNunitarydet}), then the recursion coefficients are given by
$R_n=\tau_{n+1}\,\tau_{n-1}/\tau_n^2$ and the free energy by $F_N={\cal F}_N$,
where ${\cal F}_n=\log(\tau_n/\tau_n^{\rm Gauss})$. Replacing sequences by
functions of $x\in[0,1]$ as prescribed by (\ref{calRreplace}), we then have to
leading order in $N$,
\beq
\log\frac{(1+x)R(x)}x=\frac{\partial^2{\cal F}(x)}{\partial x^2}~~~~~~,~~~~~~
F_0={\cal F}(1)
\label{Rspecheatrel}\eeq
The relations (\ref{Rspecheatrel}) show that the function $R(x)$ is related to
the specific heat $u_0$ of the matrix model, and also that any singularity in
the free energy will occur at the boundary $x=1$ of the domain of $R(x)$. These
facts will enable us to construct the topological expansion of the fermionic
matrix model from the recursion coefficients $R_n$ of the orthogonal
polynomials. We remark that it is possible to demonstrate from the constrained
sum over $SU(N)$ group characters of section 2.3 that the partition function is
real and has a definite value for any $N$. Combining this with our knowledge
from large $N$ loop equation calculations, wherein the system goes to a
definite limit, one can argue that there is a definitive $\frac1N$ expansion of
the free energy. The phase transition at large $N$ can thereby be understood as
a sort of percolation transition in the Young tableaux, whereby the system
evolves into a phase in which the number of filled $SU(N)$ Young tableaux grows
asymptotically.

\subsection{Main Recursion Relation}

We will now begin quantifying the above discussion. For definiteness, in the
remainder of this section we will deal primarily with the simplest non-Gaussian
model which is given by the quadratic potential
\beq
V(z)=z+\frac g2\,z^2
\label{quadpot}\eeq
although, as we will discuss, the results we obtain are universal. The
potential (\ref{quadpot}) is difficult to treat within the loop equation
approach because the endpoints of the support of the corresponding spectral
density are located asymmetrically in the complex plane \cite{acm}. Indeed, one
power of
the technique that we develop is that it is insensitive to the parity symmetry
(or lack thereof) of the potential. This will account for the universality
properties of the solution that we shall find. In this subsection we will
derive the main recursion relation corresponding to the potential
(\ref{quadpot}) that will be used to construct the topological expansion.

To solve the recursion relations (\ref{Rneqn}), (\ref{Pneqn}) and
(\ref{PQRrel}) corresponding to (\ref{quadpot}), we first use (\ref{Precrel})
and (\ref{calPQR}) to determine the coefficients
\bea
P_{n,n-1}^{[1]} & = & {\cal P}_{n-1}^{(n)}~=~
Q_{n-1} Q_n R_n ( R_{n+1} -1)\nn \\
P_{n,n-2}^{[1]} & = & {\cal P}_{n-2}^{(n)}~=~
Q_{n-2} Q_{n-1} Q_n R_{n-1} R_n ( R_{n+1} -1)
\eea
This leads to the closed set of recursion relations
\bea
(N+n) R_n - n &=& g\,N\,Q_{n-1} Q_n R_n ( R_{n+1} - 1) \label{first} \\
n P_{n-1} - (n-1) P_n& = & g\,Q_{n-2} Q_{n-1} Q_n R_{n-1} R_n  ( R_{n+1}
- 1) \label{second}\\
N( P_{n+1} - P_n)  &=&  Q_n  ( R_{n+1} -1) \label{third}
\eea
{}From the relations (\ref{second}) and (\ref{third}) one can solve for the
coefficients $P_n$ as
\beq
N\,P_n =  g\,N\,Q_{n-2} Q_{n-1} Q_n R_{n-1} R_n  ( R_{n+1}
- 1)  + n\,Q_{n-1} (R_n -1)
\label{Pnexpl}\eeq
Substituting (\ref{Pnexpl}) into (\ref{third}) and using (\ref{first}) to
reduce the degree of the $Q_n$'s in the resulting equation yields
\bea
0&=&Q_{n-1}  R_{n}\Bigl[(N+n+1) R_{n+1} - n - 1\Bigr] -
Q_{n-2}  R_{n-1} \Bigl[ (N+n) R_{n} - n\Bigr]\nn\\& &+\,n\,Q_n (R_{n+1} -1) -
n\,Q_{n-1} (R_{n} -1)
\label{qs}\eea
We now use (\ref{first}) to define the quantity
\beq
\Omega_n\{R\}\equiv Q_n Q_{n-1} =\frac1g\,
\frac{ (N+n) R_n - n}{N\,R_n(R_{n+1}-1)}
\label{Omegandef}\eeq
and multiply the relation (\ref{qs}) through by $Q_n$. Using (\ref{second})
along with (\ref{Pnexpl}), we can then solve for $Q_n$ in terms of the $R_n$'s
alone as
\beq
Q_n\{R\}^2=\Omega_n\{R\}^2\,\frac{n(R_n-1)-R_n\Bigl[(N+n+1)R_{n+1}-n-1\Bigr]}
{n\,\Omega_n\{R\}(R_{n+1}-1)-\Omega_{n-1}\{R\}\,R_{n-1}\Bigl[(N+n)R_n-n\Bigr]}
\label{Qnsol}\eeq

We may therefore write down a recursion relation involving {\it only} the
coefficients $R_n$ in the form
\beq
Q_n\{R\}^2\,Q_{n-1}\{R\}^2=\Omega_n\{R\}^2
\label{Rrecrel}
\eeq
where the function $Q_n\{R\}^2$ is given by (\ref{Qnsol}) and $\Omega_n\{R\}$
by (\ref{Omegandef}). Notice that the coupling constant $g$ completely cancels
out in this final equation which is an involved non-linear recursion relation
for the $R_n$'s solely in terms of $n$ and $N$. The dependence of $R_n$ on $g$
comes about purely as a boundary condition and can only be recovered after
solving for the dynamics of $R_n$. However, we will see that the entire
topological expansion can be constructed without ever knowing this dependence
explicitly. All information about the continuum limit will come from the
details of the main recursion relation (\ref{Rrecrel}).

\subsection{Planar Limit}

The spherical continuum limit of the matrix model is found by taking the
limit $N\to\infty$. Let us first describe the polymer contribution $f_0$ in
(\ref{F0}). For the potential (\ref{quadpot}), the solution of the quadratic
saddle-point equation (\ref{zetaeq}) which is regular at $g=0$ is given by
\beq
\zeta=\frac1{2g}\,\left(\sqrt{1+4g}-1\right)
\label{zetaquad}\eeq
so that the corresponding fermionic vector free energy (\ref{f0}) is
\beq
f_0(g)=\frac12+\frac1{4g}\,\left(\sqrt{1+4g}-1\right)-\log\frac1{2g}\,
\left(\sqrt{1+4g}-1\right)
\label{f0quad}\eeq
There is a phase transition at $g=g^{(0)}=-\frac14$ below which the free energy
becomes complex-valued. Near this critical point, (\ref{f0quad}) behaves as
$f_0(g)\sim(g-g^{(0)})^{3/2}$, which identifies the corresponding string
susceptibility exponent as $\gamma_{\rm str}^{(0)}=+\frac12$. The phase
transition is therefore associated with the evolution of the system into a pure
branched polymer phase of two-dimensional quantum gravity. In fact, the
function (\ref{f0quad}) is related to the usual bosonic vector model free
energy ${\cal F}_{\rm vec}$, which is the generating function for pure,
connected branched polymer graphs, by $f_0=1-2{\cal F}_{\rm vec}$ \cite{ss2}.
The $N$ dimensional Euclidean radial coordinate $r$ of the bosonic vector model
is related to the saddle-point (\ref{zetaquad}) by $r^2=\frac12\,\zeta$. One
can also compute the one-loop fluctuations around the saddle-point
(\ref{zetaquad}) and show that the resulting free energy is logarithmically
divergent at $g=0$ \cite{ss2}. Therefore, in the regime $g>0$ there is no
polymer behaviour and one may imagine a ``barrier'' in the large $N$ limit of
the theory at the Gaussian point $g=0$. It is further possible to carry out a
double scaling expansion of the fermionic vector model partition function
(\ref{h0expl}) in the limit
$N\to\infty,g\to g^{(0)}$ with the parameter $N(g-g^{(0)})^{3/2}$ held fixed.
The resulting expression is an alternating, Borel summable series which can be
expressed as a Bessel function \cite{ss2}.

We now consider the large $N$ behaviour associated with the second term in the
free energy (\ref{F0}). Replacing the discrete $R_n$ coefficients with the
large-$N$ continuous function $R(x)$ as prescribed by (\ref{calRreplace}),
after some algebra we obtain from the recursion relation (\ref{Rrecrel}), at
leading order in $N$, the first order non-linear ordinary differential equation
\beq
\frac{dR}{dx}=\frac{R\,\Bigl(R-1\Bigr)^2\,\Bigl(x+2x\,R-3(1+x)\,R^2\Bigr)}
{x^2 - 2 x^2\,R +4 x (1+x)\,R^2 -2 ( 1+ x) (2+3 x)\,R^3 + 3 ( 1+x)^2\,R^4}
\label{spherediff}
\eeq
The dependence of $R$ on the coupling constant $g$ arises only from the constant
of integration of (\ref{spherediff}). It is known from the general theory of
ordinary differential equations \cite{odes} that the solutions $R(x)$ to
equations such as (\ref{spherediff}) possess algebraic non-analytic behaviour.
Since $dR/dx$ in (\ref{spherediff}) is given as the quotient of a quintic
polynomial in $R$ by a quartic one, it follows that the function $R(x)$ is
finite for all finite values of $x$ \cite{odes}. The only singularities which
can arise from this differential equation occur when the denominator on the
right-hand side vanishes and the numerator is non-zero. In light of the
discussion of the previous subsection, we will demand that these singular
points occur at the boundary $x=1$ of the domain of the function $R(x)$. At
$x=1$, the denominator of the right-hand side of (\ref{spherediff}) has four
distinct zeros $R_c$ determined as the solutions of the quartic equation
\beq
1 - 2 R_c +8 R_c^2 -20 R_c^3 + 12 R_c^4=0
\label{rceqn}
\eeq

Denoting the corresponding critical value of the coupling constant $g$ by
$g_c$, we make an ansatz for the form of the function $R(x;g)$ near the
critical point,
\beq
R(x;g) = R_c +a\cdot(g_c - g x)^{-\gamma_{\rm str}} + \dots
\label{Ransatz}\eeq
where $a$ is some constant and the ellipsis denotes terms which are less
singular at the critical point. This ansatz ensures that for $g\sim g_c$, the
singularities occur at $x=1$, or alternatively that at $x=1$ the free energy
${\cal F}(x)$ becomes non-analytic at $g=g_c$. Substituting (\ref{Ransatz})
into the differential equation (\ref{spherediff}), we find for each branch
$R_c$ of the equation (\ref{rceqn}) the leading behaviour
\beq
\frac{d R}{dx} \sim \frac{1}{ R_c - R(x)}
\eeq
Comparing with (\ref{Ransatz}) fixes the exponent $\gamma_{\rm str}=-\frac12$,
and so the critical point $g=g_c$ of the fermionic matrix model describes a
continuum limit that lies in the same universality class as pure
two-dimensional quantum gravity in the planar limit. Note that this criticality
argument requires no knowledge of the precise value of the critical coupling
$g_c$. The same will be true of the double scaling limit which will be analysed
in the next subsection. It is possible to show using loop equations \cite{ss1}
that $g_c>0$. Thus the continuum surface behaviour occurs on the opposite side
of the Gaussian point relative to the branched polymer phase transition, and
the two phases cannot be connected together in a smooth way.

We conclude our analysis of the planar limit of the adjoint fermion one-matrix
model by briefly describing how it extends to more general
potentials. For example, consider a potential of the generic form $V(z)=z+\frac
gK\,z^K$, $K\geq2$. The recursion equations (\ref{Rneqn}) and (\ref{Pneqn}) in
this case require knowledge of the coefficients $P_{n,n-1}^{[K]}$ and
$P_{n,n-2}^{[K]}$ determined from (\ref{Precrel}) and (\ref{calPQR}). It is
straightforward to see from these relations that $P_{n,n-1}^{[K]}$ is of degree
$K$ in both the $Q_n$'s and the $R_n$'s, while $P_{n,n-2}^{[K]}$ is of degree
$K+1$. In the large $N$ limit, the recursion equations will therefore assume
the form
\bea
(1+x)R-x&=&g\,Q^K\,W_1[R]\label{genfirst}\\
P-x\,\frac{\partial P}{\partial x}&=&g\,Q^{K+1}\,W_2[R]\label{gensecond}\\
\frac{\partial P}{\partial x}&=&Q\Bigl(R-1\Bigr)
\label{genthird}\eea
where $W_1[R]$ and $W_2[R]$ are polynomials of degree $K$ and $K+1$ in the
function $R$, respectively, which each contain the factor $R-1$. Using
(\ref{genfirst}) we can solve for $Q^K$ and write (\ref{gensecond}) as an
equation determining the function $Q$. Comparing with the solution for $Q$
obtained using (\ref{genthird}), the relation (\ref{gensecond}) then becomes a
first order, linear, homogeneous differential equation for the function $P$.
Substituting the resulting solution for $P$ back into (\ref{genfirst}) yields
an integral equation for the function $R$ which, upon differentiation, can be
transformed into a first order non-linear ordinary differential equation of the
form
\beq
\frac{dR}{dx}=\frac{J_2[R;x]}{J_1[R;x]}
\label{diffeqsphericalgen}\eeq
where $J_1[R;x]$ and $J_2[R;x]$ are polynomial functions of their arguments of
respective degrees $2K$ and $2K+1$ in $R$ which are independent of the coupling
constant $g$. Arguing as we did above, the differential equation
(\ref{diffeqsphericalgen}) will admit critical behaviour for the function $R$
of the square root type, i.e. that of pure gravity. Provided that the zeros of
the function
$J_1[R;x]$ are non-degenerate at $x=1$, this will be the only type of
critical behaviour that the model admits. It is of course natural that pure
gravity exists for a generic matrix potential. Multicritical points require
more complicated potentials with two or more coupling constants and an
appropriate fine-tuning of them. In this case the differential equations
derived analogously to those above become highly non-linear and quite involved.
But the general picture will be the same, namely the function $R$ will be
determined by some differential equation whose polynomial coefficients (in $R$
and $x$) can be tuned to obtain higher order critical points. In this way we
can recover, in the planar limit, the standard universality classes of
conformal matter coupled to two-dimensional quantum gravity \cite{fgz}.

\subsection{Double Scaling Limit}

In this subsection we will study the all genus expansion of the specific heat
\beq
u(g,N)=\sum_{h=0}^\infty N^{-2h}\,u_h(g)~~~~~~,~~~~~~u_h(g)=F_h''(g)
\label{stringsuscexp}\eeq
where $F_h(g)$ is the genus $h$ contribution to the free energy $F_N(g)$ and
the genus $h$ susceptibility is a homogeneous function of degree $-2h$,
$u_h(a\cdot g)=a^{-2h}\cdot u_h(g)$. From the analysis of the previous
subsection, we know that the leading singular behaviour of
(\ref{stringsuscexp}) at genus zero is $u_0(g)\sim\sqrt{g-g_c}$. In order to
keep the partition function $Z_0(g)\sim\e^{N^2F_0(g)}$ finite in the large $N$
limit, we should
therefore perform the double scaling limit $g\to g_c,N\to\infty$ by keeping
fixed the parameter
\beq
\kappa = N^{4/5}\,( g_c- g)
\label{kappadef}\eeq
We now define a scaling function $u(\kappa)$ which captures the contributions
to the string susceptibility from all genera in the double scaling limit
$g\to g_c,N\to\infty$ with the variable (\ref{kappadef}) fixed. For this, we
introduce the large-$N$ expansion parameter $\epsilon=\frac1N$ and write
the total susceptibility (\ref{stringsuscexp}) in the vicinity of the critical
point as
\beq
u(g,N)=(\epsilon^2)^{4/5}\,a_0+\epsilon^2\,
u(\kappa)+(\epsilon^2)^{6/5}\,a_1\,\kappa+O\left((\epsilon^2)^{7/5}
,\kappa^2\right)
\label{stringsuscallhdef}\eeq
where $a_n$ are constants. By defining the new effective variable
\beq
\xi=\epsilon^{-4/5}\,( g_c - g x)
\label{xidef}\eeq
in the large $N$ limit, this scaling function may be determined from the
$\xi=\kappa$ limit of a scaling ansatz for the function $R(x)$ in terms of an
unknown function $u(\xi)$,
\beq
R(x) = R_c +\epsilon^{2/5}\,u(\xi)
\label{DSansatz}\eeq
where $R_c$ is a solution of (\ref{rceqn}). By substituting (\ref{DSansatz})
into (\ref{F0}) and changing variables from $x$ to $\xi$ in the integral, we
find that, up to irrelevant constants (which may be absorbed by suitable
rescaling using the homogeneity of the specific heat) and terms which
vanish as $\epsilon\to0$, the free energy in the double scaling limit is
determined as
\beq
F(\kappa)=f_0+\int\limits_{\epsilon^{-4/5}\,g_c}^\kappa d\xi~(\kappa-\xi)
\,u(\xi)
\label{FDS}\eeq
It follows that
\beq
u(\kappa)=F''(\kappa)
\label{uFrel}\eeq
and so the problem of obtaining the topological expansion of the fermionic
matrix model is thereby reduced to the task of finding the solution of the main
recursion relation (\ref{Rrecrel}) for the function $R(x)$ in this special
limit. Note that the polymer free energy (\ref{f0quad}) contributes only an
irrelevant constant in the double scaling limit about the positive-valued
critical point $g_c$.

We now rewrite (\ref{Rrecrel}) using (\ref{calRreplace}) and take the
limit of large-$N$ while holding the quantity $\kappa$ fixed through the
relation $g= g_c-\kappa N^{-4/5}$. We then substitute in the scaling ansatz
(\ref{DSansatz}) and rewrite derivatives according to the change of variables
(\ref{xidef}), i.e. $\partial_x\sim-\epsilon^{-4/5}\,\partial_\xi$. After some
algebra, we find, as the coefficient of the leading order $\epsilon^{4/5}$
term, a non-linear differential equation for the specific heat $u(\xi)$,
\bea
0 & = & 6 g_c\left( 3 - 22R_c + 80R_c^2 - 192R_c^3 + 376R_c^4 -
632R_c^5  + 752R_c^6 - 512R_c^7 + 144R_c^8\right)\nn\\& &\times\,
u(\xi)\,u'(\xi)+\,R_c\left( 1 - 5R_c + 10R_c^2 - 10R_c^3 + 4R_c^4\right)\nn \\
&& \times\,\left[6R_c\Bigl(R_c -1\Bigr)^2\left( 6R_c^2 - 2R_c -1\right) -
g_c^3\,\left( 1 - 2R_c + 20R_c^2 - 56R_c^3 + 36R_c^4
\right)\,u'''(\xi)\right]\nn\\& &
\label{nonlinDE}\eea
We now integrate (\ref{nonlinDE}) up once and use an inessential shift of the
independent variable $\xi$ to eliminate the constant term. After applying
(\ref{rceqn}) and a rescaling
using the homogeneity of the function $u(\xi)$ (which preserves the
relationship (\ref{uFrel})), we arrive at the parameter-free equation
\beq
u''(\xi) + u(\xi)^2 =\xi
\label{paineqn}\eeq
The non-linear differential equation (\ref{paineqn}) governs the behaviour of
the partition function of the fermionic random surface model to all orders in
the genus expansion, with the parameter $\xi^{5/4}$ identified as the
renormalized string coupling constant. We note again the remarkable feature
that one never needs to know the precise location of the critical point to
arrive at this
equation. In the present case the critical coupling constant $g_c$ can be
consistently rescaled out of the pertinent equations. This parametric
independence is indicative of the universality of the double scaling equation
(\ref{paineqn}) for the given class of generic polynomial potentials.

The differential equation (\ref{paineqn}) is known as the Painlev\'{e} I
equation and it is solved by the first Painlev\'{e} transcendent. This equation
differs from the versions which usually appear in matrix models \cite{fgz} by a
change in sign of the specific heat $u\to-u$. The boundary conditions which
$u(\xi)$ satisfies
follow from the analysis of the planar limit of the previous subsection. The
genus zero contribution arises in the limit of large, positive $\xi$ where the
leading behaviour is $u(\xi)^2 =\xi$. As a consequence, we are led to postulate
an asymptotic expansion for $u(\xi)$ of the form
\beq
u(\xi) = \sqrt{\xi}\,\left( 1+ \sum_{k=1}^\infty u_k\,\xi^{-5k/2} \right)
\label{sersoln}
\eeq
By substituting (\ref{sersoln}) into (\ref{paineqn}), the first few
coefficients are easily calculated to be
\bea
u_1&=&\frac{1}{8}\nn\\u_2&=&-\frac{49}{128}\nn\\u_3&=&\frac{1225}{256}
\nn\\u_4&=&-\frac{4412401}{32768}\nn\\u_5&=&\frac{220680075}{32768}\nn\\
u_6&=&-\frac{2207064977649}{4194304}\label{1stcoeffs}\\&\vdots&\nn
\eea
and in general they are determined by the recursive equation
\beq
u_k=-\frac{(5k-6)(5k-4)}8\,u_{k-1}-\frac12\,\sum_{n=1}^{k-1}u_k\,u_{k-n}
{}~~~~~~,~~~~~~k\geq2
\label{ukreceq}\eeq
With the normalization of the spherical specific heat taken as $u_0=1$, we see
from (\ref{1stcoeffs}) that the coefficients of the genus expansion have {\it
precisely} the same absolute numerical values as those obtained for pure
two-dimensional quantum gravity, i.e. from the usual Painlev\'e expansion
\cite{fgz,2Dbook}. In particular, they have the same high rate asymptotic
growth in magnitude. However, the most notable feature of the coefficients
(\ref{1stcoeffs}) is the fact that they oscillate in sign (In the Hermitian
cases, all $u_k$ are negative). This raises the possibility that the asymptotic
series solution (\ref{sersoln}) may in fact be Borel summable. If so, then the
free energy would give an unambiguous definition of the genus expansion of pure
``fermionic'' gravity in powers of
$N^{-2} \sim\xi^{-5/2}$. A numerical solution of the equation (\ref{paineqn})
is possible and the result is shown in fig.~1. This confirms numerically that
the free energy exists as a well-defined function and is real-valued on the
positive real $\xi$-axis. In the next subsection we shall prove that the
coefficients $u_k$ of the asymptotic series (\ref{sersoln}) alternate in sign
to arbitrarily large orders of the genus expansion, and, moreover, that there
is a well-defined Borel resummation of $u(\xi)$ for $\xi>0$.

\begin{figure}[htb]
\epsfxsize=3in
\bigskip
\centerline{\epsffile{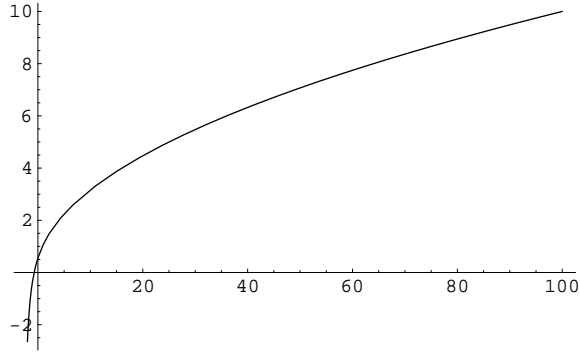}}
\caption{\baselineskip=12pt{\it Plot of the solution $u(\xi)$ versus $\xi$ of
the Painlev\'e equation with the asymptotic boundary condition $u(\xi)^2=\xi$
for $\xi\to+\infty$. The first double pole singularity (corresponding to a zero
of the partition function) occurs on the negative axis at $\xi\simeq-3.1477$.}}
\bigskip
\end{figure}

\subsection{Borel Summability}

In this subsection we will argue analytically, along the lines of \cite{fgz},
that the asymptotic series (\ref{sersoln}) defines a unique function $u(\xi)$.
For this, we consider the Borel transform
\beq
B(s) = \sum_{k=1}^\infty \frac{u_k}{( \beta k)!}\,s^k
\label{boreltform}\eeq
where the constant $\beta$ will be self-consistently determined by the
condition that this series has a finite radius of convergence. A solution of
the original Painlev\'{e} equation (\ref{paineqn}) is then given by
\beq
u(\xi) = \sqrt{\xi}\,\left( 1 - \int\limits_0^\infty dt~\e^{-t}
\,B\left(t^\beta\,\xi^{-5/2}\right)
\right)
\label{borelint}\eeq
The crucial issue now is whether or not the integral in (\ref{borelint})
actually exists. We will show that there is a contour running through the ${\rm
Re}\,t>0$ region of the complex $t$-plane from $t=0$ to $t=\infty$ along which
the integral transform (\ref{borelint}) converges. We will thereby argue
that the Borel resummation (\ref{borelint}) of the specific heat is
well-defined and real-valued.

The function (\ref{boreltform}) has singularities in the complex $s$-plane with
branch cuts running between them and infinity. If some of these singularities
lie on the positive $s$-axis then the integral definition (\ref{borelint}) is
ambiguous. We will now extract the large order behaviour of the coefficients
$u_k$ and argue that the singularities of the Borel transform $B(s)$ all lie on
the negative $s$-axis. From (\ref{boreltform}) and (\ref{borelint}) it follows
that the coefficients of the asymptotic expansion (\ref{sersoln}) are given by
\beq
u_k = \frac{1}{2 \pi i}\,\int\limits_0^\infty dt ~ \e^{-t}\,\oint
\limits_{s=0}\frac{ds }{s^{ k+1}}~B\left( t^\beta\,s\right)
\label{ukBs}\eeq
The contour of integration in (\ref{ukBs}) can be extended out to infinity
around the cuts of the Borel transform in the complex $s$-plane. In the limit
$k\to\infty$, the contour integral in (\ref{ukBs}) is dominated by
contributions along the branch cut of $B(s)$ in the complex $s$-plane which
begins at the point $s=s_0$ that is closest to the origin. We then have, for
large $k$,
\beq
u_k\sim\int\limits_0^\infty dt ~ \e^{-t}\,\int
\limits_{s_0/t^\beta}^\infty\frac{ds }{s^{ k+1}}~{\rm Disc}\,B
\left( t^\beta\,s\right)\sim\int\limits_{0}^\infty
\frac{ds }{s^{ k+1}}~\int\limits_{(s_0/s)^{1/\beta}}^\infty dt ~ \e^{-t}
{}~{\rm Disc}\,B\left( t^\beta\,s\right)
\label{ukestimate}\eeq
where we have defined the discontinuity of the Borel transform
\beq
{\rm Disc}\,B(s)=B_+(s)-B_-(s)~~~~~~,~~~~~~B_\pm(s)=B(s\pm i0)
\label{Discdef}\eeq
for $s$ a point on its cut.

Going back to (\ref{borelint}), we see from (\ref{ukestimate}) that we should
study the functions $u_\pm(\xi)$ corresponding to the Borel transforms on
either side of the dominant cut,
\beq
u_\pm(\xi) = \int\limits_{(s_0\,\xi^{5/2})^{1/\beta}}^\infty dt ~ \e^{-t}
\,B_{\pm}\left( t^\beta\,\xi^{-5/2}\right)
\label{upmdef}\eeq
Since each of the functions (\ref{upmdef}) solves the Painlev\'e equation
(\ref{paineqn}), it is convenient to define their sum and difference
\beq
u_{\rm d}(\xi)= u_+(\xi)- u_-(\xi)= {\rm Disc}\,B(\xi)~~~~~~,~~~~~~
u_{\rm s}(\xi)=\mbox{$\frac12$}\,\Bigl(u_+(\xi)+ u_-(\xi)\Bigr)
\label{udusdefs}\eeq
which on using (\ref{paineqn}) are seen to obey the coupled system of
differential equations
\bea
u_{\rm d}''(\xi)+ 2 u_{\rm d}(\xi)\,u_{\rm s}(\xi)& = & 0\label{Bfirst}\\
u_{\rm d}''(\xi)+ u_{\rm s}(\xi)^2 + \frac{1}{4}\,u_{\rm d}(\xi)^2 & = &\xi
\label{Bsecond}\eea
These equations can be solved in the WKB approximation.
In the planar limit $\xi\to\infty$, we self-consistently assume that $u_{\rm
d}(\xi)$ is of sub-leading order in (\ref{Bsecond}) and thereby find, to
leading order, the solution $u_{\rm s}(\xi)\sim\sqrt{\xi}$. By substituting
this into (\ref{Bfirst}), to leading order we have the first order linear
differential equation for the function $u_{\rm d}(\xi)$,
\beq
u_{\rm d}''(\xi) + 2\sqrt{\xi}\,u_{\rm d}(\xi)=0
\label{udeqn}\eeq
which can be solved in terms of Bessel functions as $u_{\rm
d}(\xi)=\sqrt{\xi}\,Z_{\frac25}\left(\frac{4\sqrt2}5\,\xi^{5/4}\right)$. The
solution for $\xi\to\infty$ is therefore given by
\beq
\frac{u_{\rm d}(\xi)}{\sqrt{\xi}} \sim\xi^{-5/8}\,\cos{\left( \frac{4
\sqrt{2}}{5}\,\xi^{5/4} \mp\frac{5 \pi}{4} - \frac{\pi}{4} \right)}
\label{udasympt}\eeq

We are finally ready to estimate the large order behaviour of the coefficients
$u_k$. For this, it is convenient to change variables in order to bring the
asymptotic expansion for $\xi\to\infty$ to $\omega\to0$ through the definition
$\omega =\xi^{-5/4}$. Then, from (\ref{ukestimate})--(\ref{udusdefs}) and
(\ref{udasympt}), we find as $k \rightarrow \infty$,
\beq
u_k  \sim \int \frac{d \omega}{\omega^{2 k +1}}~\omega ^{2/5}\,u_{\rm d}
\left(\omega^{-4/5}\right)\sim
\int \frac{d \omega}{\omega^{2 k +1} }~\sqrt\omega\,
\cos{\left( \frac{4 \sqrt{2}}{5 \omega} \mp
\frac{5 \pi}{4} - \frac{\pi}{4} \right)}
\label{ukestasympt}\eeq
For each choice of sign the integral (\ref{ukestasympt}) gives, up to constant
factors, the result
\beq
u_k\sim\left(-\frac{32}{25}\right)^{-k}\,\Gamma
\left(2k-\mbox{$\frac12$}\right)~~~~~~{\rm for}~~k\to\infty
\label{uasympt}\eeq
The asymptotic estimate (\ref{uasympt}) contains a large amount of information.
 Going back to the Borel transform (\ref{boreltform}), the large order
behaviour of $u_k$ fixes $\beta=2$ and gives
\beq
B(s)\sim \sum_k\left( -\frac{32}{25} \right)^{-  k}\,s^k
\label{borelexpl}\eeq
We see therefore that the Borel transform $B(s)$ is well-defined in an open
region of the complex $s$-plane and, moreover, that its first singularity
appears on the negative real $s$-axis. Assuming that all of its singularities
are so restricted, we can expect that the asymptotic series (\ref{sersoln}) can
be Borel resummed through the integration in (\ref{borelint}) to define a
unique real function $u(\xi)$ on the positive real $\xi$-axis. The results of a
numerical analysis (fig.~1) support these arguments.

{}From this analysis it follows that the double-scaled partition function of
the adjoint fermion one-matrix model leads to a completely well-defined genus
expansion of the corresponding random surface theory. The coefficients
(\ref{uasympt}) have precisely the same large order behaviour
$\Gamma(2k-\frac12)$ as those of pure two-dimensional quantum gravity. In fact,
the topological expansion of the fermionic matrix model is identical term by
term to that of the usual Hermitian matrix models. However, the alternating
nature of the asymptotic series expansion allows one to define the string
susceptibility in an unambiguous way through its Borel resummation
(\ref{borelint}), at least for positive values of the string coupling $\xi$. In
this way we may think of the fermionic matrix model as lending a well-defined
version of the generating functions provided by Hermitian matrix models. It
gives an analytic solution to the problem of counting fermionic random
triangulations on arbitrary genus Riemann surfaces, whose asymptotic expansion
is Borel summable but otherwise coincides with the usual Painlev\'e expansion
of two-dimensional quantum gravity.

The usual movable, double pole singularities of the Painlev\'e I equation
appear as well in the present case, but this time they appear for {\it
negative} values of the cosmological constant $\xi$. The existence of such
poles is inconsistent with the loop equations of two-dimensional quantum
gravity \cite{david}. Pole-free solutions of Painlev\'e equations are provided
by the
triply-truncated Boutroux solutions, but these are complex-valued and do not
lead to physically acceptable free energy functions. However, in the present
situation there is a unique, well-defined real-valued specific heat on the
positive $\xi$-axis\footnote{We thank
A.A. Kapaev for pointing out the relevant mathematical literature \cite{kapaev} on this
point.} which should have a meaningful analytical continuation to
$\xi<0$. It provides an unambiguous definition of a whole branch of the random
surface theory, and, given the relation between the double scaling variable
$\xi$ and the continuum cosmological constant, the continuum limit of the
discretized model is indeed well-defined. This suggests that by restricting our
``fermionic gravity'' model to $\xi>0$, we could define a model of
two-dimensional quantum gravity with the canonical characteristics order by
order in the genus expansion, but whose full asymptotic series makes perfect
sense and provides a well-defined non-perturbative definition of the theory.

In the next and final section we will argue that this is indeed the case. In
the usual Hermitian one-matrix models, the problem with the double scaling
limit, in which the original matrix integral can only be defined by an
analytical continuation, can be traced back to an instability in the
corresponding double-scaled eigenvalue model. The non-compactness of the
double-scaled eigenvalue space $\IR$ and the form of the effective potential
for the eigenvalues demonstrates that the definition of the critical points of
the model is unstable to the tunneling of eigenvalues into a different
configuration \cite{davidinst}. This leads to instanton solutions in the
single-well eigenvalue
model, and thereby explains the complexity of the free energy. However, in
the present situation the Grassmann matrix integral is completely well-defined
at finite $N$, in contrast to the Hermitian cases, and there is no reason {\it
a priori} to expect that the model becomes unstable in any way at large $N$.
Indeed, with the mapping of the fermionic matrix model onto a unitary matrix
model, we see that the matrix integral is determined by an eigenvalue space
whose topology is that of a circle. Being a compact space, there is no
asymptotic behaviour in the effective eigenvalue action, and thus the
compactness could have the effect of eliminating the
eigenvalue tunneling problem. We see then that the unitary representation of
the fermionic matrix model provides, at least naively, an eigenvalue
description of the random surface model which reflects the reasons why the
matrix integral is superior to its Hermitian counterparts. We shall tackle this
problem using an operator theoretic approach which gives the fermionic analog
of the (generalized) KdV hierarchical structure of the Hermitian one-matrix
models \cite{fgz,aaaaa}. The integrable flows that we shall find are similar to
those of the bosonic models, but with some very important changes. The most
important difference will be the absence of a translational symmetry in
eigenvalue space. This feature can be interpreted as allowing one to restrict
the double-scaled specific heat to positive values of the cosmological constant
$\xi$. It may then follow that the partition function of the fermionic matrix
model serves as an unambiguous definition of that for two-dimensional quantum
gravity.

We stress that the resulting Borel summability of the fermionic case is very
different from that of Borel summable bosonic models, such as the conformal
field theory of the Yang-Lee edge singularity in which the lattice system
couples to an imaginary magnetic field and is therefore non-unitary \cite{fgz}.
The coefficients of the genus expansion of the adjoint fermion one-matrix model
are all real-valued\footnote{In \cite{matone} a similar, but non-unitary, alternating 
genus expansion was used to formulate a model of two-dimensional quantum gravity
in terms of integrals over the moduli spaces of punctured spheres.}. 
The stability of the fermionic model is also quite
distinct from the stabilizations provided by supersymmetric and stochastic
quantization methods \cite {mp}, because these latter models violate the KdV
flow structure of two-dimensional quantum gravity at a non-perturbative level
\cite{stochflow}. In the present case, we will see that the appropriate
integrable hierarchical structure is a deformation of that for gravity, and so
the KdV flow structure continues to hold in a certain sense. The precise
meaning of this deformation comes from the
worldsheet interpretation of the fermionic one-matrix model. This is the model
of fermionic gravity that we alluded to in section 2.1, wherein ``Penner matter
states'' are located at the vertices of the corresponding surface
discretizations. 

\section{Operator Formalism}
\setcounter{equation}{0}

The discussion at the end of the previous section motivates the development of
an operator approach to analyse more carefully the properties of the adjoint
fermion one-matrix model within the orthogonal polynomial formalism. The main
purpose is two-fold. First of all, it will allow us to relate the partition
function of the fermionic matrix model to the $\tau$-function of an integrable
hierarchy \cite{aaaaa,tau,inthi}. The usefulness of this association is that it
partitions the solutions of the fermionic matrix model into universality
classes which are related to
well-known integrable hierarchies, and thereby formulates the partition
function in an invariant way as the solution to a set of differential
equations, in particular in the double scaling limit. Secondly, given this
relationship, we will obtain a clear geometric picture of the origin of the
alternating, Borel summable genus expansion of the model. This will allow us to
clarify the description of the string susceptibility as a well-defined,
non-perturbative generating function for the fermionic gravity model, as
alluded to at the end of the previous section, and at the same time it will
present the appropriate generalization of the KdV flow structure that
characterizes the gravity model, lending a complementary characterization to
the worldsheet description. In the following we will begin by introducing the
formalism and deriving the main equations that will be required. We will then
identify the integrable hierarchy of which the fermionic partition function is
a $\tau$-function, and derive the Virasoro constraints which must be satisfied
by the flows.

\subsection{String Equations}

We begin by introducing multiplication operators $\bf Q$ and $\M$ defined on
the space $\IC[z]\otimes\IC[\bar z]$ by
\beq
({\bf Q}\,\Phi_m)(z)\equiv
z\,\Phi_m(z)~~~~~~,~~~~~~(\M\,\Lambda_m)(z)\equiv\frac1z\,\Lambda_m(z)
\label{QMdef}\eeq
and differentiation operators $\bf P$ and $\L$ by
\beq
({\bf P}\,\Phi_m)(z)\equiv\Phi_m'(z)~~~~~~,~~~~~~(\L\,\Lambda_m)(z)\equiv
z^2\,\Lambda_m'(z)
\label{PLdef}\eeq
The operators $\Q$ and $\M$ are related to each other in a very simple way.
{}From (\ref{QMdef}) it follows that
\beq
\Bigl\la\Q\,\Phi_n(z)\Bigm|\M\,\Lambda_m(z)\Bigr\ra=
\Bigl\la\Phi_n(z)\Bigm|\Lambda_m(z)\Bigr\ra
\label{Qunitary}\eeq
and so $\Q$ is an invertible operator with inverse
\beq
\Q^{-1}=\M^\dagger
\label{QinvbarQ}\eeq
where the adjoint ${\bf O}^\dagger$ of any operator ${\bf O}$ on
$\IC[z]\otimes\IC[\bar z]$ is defined by
\beq
\Bigl\langle F(z)\Bigm|({\bf
O}\,G)(\bar z)\Bigr\rangle=\Bigl\langle({\bf O}^\dagger\,
F)(z)\Bigm|G(\bar z)\Bigr\rangle
\label{adjointdef}\eeq

By writing $[z\,\Phi_m(z)]'$ in two different ways as
$[(\Q\,\Phi_m)(z)]'$ and $\Phi_m(z)+z\,\Phi_m'(z)$, we can infer the
canonical commutation relation
\beq
\Bigl[\P\,,\,\Q\Bigr]={\bf1}
\label{cancommrel}\eeq
Similarly, by considering $z^2\,[\frac1z\,\Lambda_m(z)]'$ we arrive at
$\left[\M\,,\,\L\right]={\bf1}$, which with the unitarity condition
(\ref{QinvbarQ}) implies the canonical commutator
\beq
\Bigl[\L^\dagger\,,\,\Q^{-1}\Bigr]={\bf1}
\label{noncanrel}\eeq
Geometrically, the operator $\P$ is the canonical conjugate of the operator
generating clockwise unit shifts on the circle, while $\L$  serves as the
canonical conjugate of that which generates counterclockwise shifts.
The operators introduced above are not all independent, but are related to
each other through a Schwinger-Dyson equation. This follows from the inner
product
\bea
\Bigl\la(\P\,\Phi_n)(z)\Bigm|\Lambda_m(z)\Bigr\ra&=&
\Bigl\la\Phi_n'(z)\Bigm|\Lambda_m(z)\Bigr\ra\nn\\&=&
(N+1)\Bigl\la\Phi_n(z)\Bigm|\mbox{$\frac1z$}\,\Lambda_m(z)\Bigr\ra
-N\Bigl\la V'(z)\,\Phi_n(z)\Bigm|\Lambda_m(z)\Bigr\ra\nn\\& &
-\,\Bigl\la\Phi_n(z)\Bigm|\Lambda_m'(z)\Bigr\ra
\label{eqmotionderiv}\eea
which implies the operator identity
\beq
\P+\L^\dagger\,\Q^{-2}=(N+1)\,\Q^{-1}-N\,V'(\Q)
\label{opid1}\eeq
The equations (\ref{cancommrel}), (\ref{noncanrel}) and (\ref{opid1}) define
the string equations of the adjoint fermion one-matrix model.

Let us now describe some basic properties of the operators introduced above.
For this, we first rewrite their actions on the {\it same} basis $\Phi_n$ of
the space of polynomials in a single variable on the circle. We have
\bea
({\bf Q}^{\pm1}\,\Phi_n)(z)&=&\sum_{m\geq0}\left[{\bf
Q}^{\pm1}\right]_{nm}\,\Phi_m(z)\nn\\({\bf
P}\,\Phi_n)(z)&=&\sum_{m\geq0}\Bigl[{\bf
P}\Bigr]_{nm}\,\Phi_m(z)\nn\\(\L^\dagger\,\Phi_n)(z)
&=&\sum_{m\geq0}\left[\L\right]_{nm}\,\Phi_m(z)
\label{opactsame}\eea
We denote by ${\bf O}_+$ the upper triangular part, including the main
diagonal, of the discrete operator $\bf O$ when represented in the basis of
polynomials in (\ref{opactsame}). Then ${\bf O}_-={\bf O}-{\bf O}_+$ is the
lower triangular part of ${\bf O}$.

{}From the definition (\ref{QMdef}) it follows that the operator $\Q$ defines a
Jacobi matrix, because
\beq
[\Q]_{nm}=0~~~~~~{\rm for}~~m-n>1
\label{JacobiQ}\eeq
In particular, its pure upper triangular part is given by
\beq
\Q_+-\sum_{n\geq0}[\Q]_{nn}\,{\bf E}_{n,n}=\sum_{n\geq0}{\bf E}_{n,n+1}
\label{QE}\eeq
where ${\bf E}_{n,m}$ are the step operators with matrix elements $[{\bf
E}_{n,m}]_{k\ell}=\delta_{kn}\,\delta_{\ell m}$. The remaining matrix elements
of $\Q$ are given by $[\Q]_{nm}={\cal P}_m^{(n)}$ in (\ref{calPQR}), along with
the recurrence relations derived in section~3. Similarly, the operator
$\Q^{-1}$ defines a Jacobi matrix in the basis (\ref{opactsame}) because
\beq
\left[\Q^{-1}\right]_{nm}=0~~~~~~{\rm for}~~n-m>1
\label{JacobiM}\eeq
Since
\beq
h_{n-1}\left[\Q^{-1}\right]_{n,n-1}=\Bigl\langle(\M^\dagger\,\Phi_n)(z)\Bigm|
\Lambda_{n-1}(z)\Bigr\rangle=\Bigl\langle\Phi_n(z)\Bigm|
\mbox{$\frac1z$}\,\Lambda_{n-1}(z)\Bigr\rangle=h_n
\label{MRnrel}\eeq
it follows that its lower triangular part is given by
\beq
\Q^{-1}_-=\sum_{n\geq1}R_n\,{\bf E}_{n,n-1}
\label{ME}\eeq
The remaining matrix elements of $\Q^{-1}$ are given by $[\Q^{-1}]_{nm}={\cal
L}_n^{(m)}$ in (\ref{calLexpl}). Furthermore, from the definition (\ref{PLdef})
we see that $\P=\P_-$ is purely lower triangular with
\beq
[\P]_{n,n-1}=n
\label{Pnn}\eeq
and since
\beq
h_{n+1}\left[\L\right]_{n,n+1}=\Bigl\langle(\L^\dagger\,\Phi_n)(z)\Bigm|
\Lambda_{n+1}(z)\Bigr\rangle=\Bigl\langle\Phi_n(z)\Bigm|
z^2\,\Lambda'_{n+1}(z)\Bigr\rangle=-(n+1)h_n
\label{Lhnrel}\eeq
it follows that $\L^\dagger=\L_+^\dagger$ is purely upper triangular,
$\left[\L\right]_{nn}=0$, with
\beq
\left[\L\right]_{n,n+1}=-\frac{n+1}{R_{n+1}}
\label{Lnn}\eeq
Note that, in this basis, the $(n,n-1)$ matrix element of the Schwinger-Dyson
equation (\ref{opid1}) coincides with (\ref{Rneqn}). This follows from the
structure of the matrix elements
\beq
h_{n-1}\left[\L^\dagger\,\Q^{-2}\right]_{n,n-1}=\Bigl\la\Phi_n(z)\Bigm|
\Lambda_{n-1}'(z)\Bigr\ra=-(n-1)h_n
\label{barPQinv}\eeq
which implies that $\left[\L^\dagger\,\Q^{-2}\right]_{n,n-1}=-(n-1)R_n$.

\subsection{Flow Equations}

We will now derive a system of flow equations for the operators above
and the partition function (\ref{rec}), which will enable us to identify the
particular integrable hierarchy at work here. We shall describe the evolutions
with respect to the discrete set of ``time'' variables $t_k\equiv
Ng_k/k$ of the generic potential (\ref{pot}). Consider the relation
\beq
\frac\partial{\partial
t_k}\Bigl\langle\Phi_n(z)\Bigm|\Lambda_{n-m}(z)\Bigr\rangle
{}~=~0~=~\Bigl\langle z^k\,\Phi_n(z)\Bigm|
\Lambda_{n-m}(z)\Bigr\rangle+\Bigl\langle
\mbox{$\frac\partial{\partial
t_k}$}\,\Phi_n(z)\Bigm|\Lambda_{n-m}(z)\Bigr\rangle
\label{diffrel1}\eeq
where $1\leq m\leq n$ and we have used the fact that
$\frac{\partial\Lambda_{n-m}(z)}{\partial t_k}$ is a polynomial of degree
at most $n-m-1$. This relation leads immediately to the evolution equation
\beq
\frac{\partial\Phi_n(z)}{\partial
t_k}=-\sum_{m=0}^{n-1}\left[\Q^k\right]_{nm}\,\Phi_m(z)
=-\left(\Q_-^k\,\Phi_n\right)(z)
\label{Phiflow}\eeq
By taking time derivatives of the equations (\ref{opactsame}), we then arrive
at the discrete operator flow equations
\bea
\frac{\partial\Q}{\partial
t_k}&=&\left[\Q\,,\,\Q_-^k\right]\label{Qflow}\\\frac{\partial\P}{\partial
t_k}&=&\left[\P\,,\,\Q_-^k\right]\label{Pflow}\\
\frac{\partial\Q^{-1}}{\partial
t_k}&=&\left[\Q^{-1}\,,\,\Q_-^k\right]\label{Mflow}\\
\frac{\partial\L^\dagger}{\partial
t_k}&=&\left[\L^\dagger\,,\,\Q_-^k\right]
\label{Lflow}\eea
The flow equations (\ref{Qflow}) constitute a discrete KP hierarchy \cite{tau}.
It is straightforward to show that these flows are
mutually commutative, so that this hierarchy is in fact integrable. This
follows from taking time derivatives of (\ref{opactsame}) and using
(\ref{Phiflow}) to get
\beq
\frac{\partial\left[\Q^k\right]_{mn}}{\partial
t_r}=\sum_{\ell=m-k}^{n-1}\left[\Q^k\right]_{m\ell}\Bigl[\Q^r\Bigr]_{\ell
n}-\sum_{\ell=m+1}^{n+k}\Bigl[\Q^r\Bigr]_{m\ell}\left[\Q^k\right]_{\ell n}
\label{Qktimederivs}\eeq
which leads to the Zakharov-Shabat equations
\beq
\frac{\partial\Q_-^m}{\partial t_n}-\frac{\partial\Q_-^n}{\partial
t_m}+\left[\Q_-^m\,,\,\Q_-^n\right]=0
\label{ZSeqns}\eeq
The matrix model actually defines a reduced, generalized KP hierarchy, because
the flow equations are to be supplemented with the constraints imposed by the
string equations.

The knowledge of the solutions $\Q$ to (\ref{Qflow}) completely
determines the partition function of the fermionic matrix model. To see this,
we consider the $(n,n-1)$ matrix element of the flow equation (\ref{Mflow}),
which using (\ref{JacobiM}) and (\ref{ME}) leads to the differential equation
\beq
\frac{\partial\log R_n}{\partial
t_k}=\left[\Q^k\right]_{nn}-\left[\Q^k\right]_{n-1,n-1}
\label{diffeq2x}\eeq
We can use (\ref{rec}) and (\ref{diffeq2x}) to obtain the flow equation for the
partition function
\beq
\frac{\partial\log Z_N}{\partial t_k}={\rm Tr}_{(N)}\left(\Q^k\right)
\label{ZNflow}\eeq
where we have defined the $N$ dimensional trace
\beq
{\rm Tr}_{(N)}({\bf O})=\sum_{n=0}^{N-1}[{\bf O}]_{nn}
\label{TrNdef}\eeq
Using the KP equation (\ref{Qflow}) and the Jacobi property (\ref{JacobiQ}) of
the $\Q$ operator, we have
\beq
\frac{\partial[\Q]_{nn}}{\partial t_k}=
\left[\Q^k\right]_{n+1,n}-\left[\Q^k\right]_{n,n-1}
\label{diffeq1x}\eeq
which, on using (\ref{diffeq2x}) with $k=1$, leads to
\beq
\frac{\partial^2\log Z_N}{\partial t_1\partial t_k}=\left[\Q^k\right]_{N,N-1}
\label{ZNflow2}\eeq
This procedure can be iterated to determine any number of time derivatives of
$\log Z_N$ in terms of the $\Q$ operators. Knowing $\Q$ therefore determines
the free energy of the matrix model up to an overall integration constant.

\subsection{Fermionic $\tau$-Function}

It is clear from (\ref{ZNflow2}) that the partition function $Z_N$ is a
$\tau$-function for the discrete KP hierarchy (\ref{Qflow}) \cite{tau}. Let us
introduce the Baker-Akhiezer functions
\beq
\Psi_n[\,\vec t\,;z]=
\frac{\e^{\frac N2\,V(z)}}{z^{\frac{N+1}2}}\,\frac{\Phi_n(z)}{\sqrt{h_n}}
\label{BAdef}\eeq
and their conjugates
\beq
\overline{\Psi}_n[\,\vec t\,;z]=
\frac{\e^{\frac N2\,V(z)}}{z^{\frac{N+1}2}}\,\frac{\Lambda_n(z)}{\sqrt{h_n}}
\label{BAconj}\eeq
which are bi-orthonormal in the standard line measure of the complex plane,
\beq
\oint\limits_{z=0}
dz~\Psi_n[\,\vec t\,;z]\,\overline{\Psi}_m[\,\vec t\,;z]=\delta_{nm}
\label{Psiortho}\eeq
The operator $\Q$ is a discrete Lax type operator which has
eigenfunctions $\Psi_n[\,\vec t\,;z]$ with eigenvalue $z$,
\beq
\Q\,\Psi_n[\,\vec t\,;z]=z\,\Psi_n[\,\vec t\,;z]
\label{eigeneqn}\eeq
The function $Z_N[\,\vec t~]$ is then the generating function for
the Lax operator $\Q$, in the sense that $\Q$ can be reconstructed from a
knowledge of $Z_N$. In fact, if we introduce the discrete $\tau$-function
$\tau_n[\,\vec t~]$ for this hierarchy by the formula (\ref{taundef})
\cite{inthi}, then from (\ref{rec}) we see that the partition function of
the fermionic matrix model is given by
\beq
Z_N[\,\vec t~]=\tau_N[\,\vec t~]
\label{ZNtau}\eeq
Formally, the $\tau$-function is a section of the determinant line
bundle over a Sato Grassmannian associated with Riemann surfaces of the
spectral parameters $z$ \cite{tau}. It may be characterized as the unique
function depending on the times $t_k$ and satisfying the given hierarchy of
constrained differential equations. It is the generating function for solutions
of the generalized KdV equations. Thus once the $\tau$-function is known,
everything about the matrix model is likewise known. The usefulness of the
relationship (\ref{ZNtau}) is that there are a large number of identities
satisfied by $\tau_N[\,\vec t~]$ that can be used to characterize the partition
function of the fermionic matrix model.

The flow equations (\ref{Qflow}) or (\ref{Qktimederivs}) constitute the Lax
representation of a discrete integrable hierarchy of differential equations of
Toda type \cite{Toda}. It is tempting therefore to characterize the fermionic
partition function as the $\tau$-function of a certain reduction of the
integrable Toda chain. However, as we will now demonstrate, this is not the
case. By equating the constant terms on both sides of the flow equation
(\ref{Phiflow}) for $k=1$ and using (\ref{calPQR}) and (\ref{Qndef}), we have
\beq
\frac{\partial p_{n,0}}{\partial t_1}=\frac{1-R_{n+1}}{l_{n+1,0}}\,
\sum_{m=0}^{n-1}p_{m,0}\,l_{m,0}\,\prod_{k=m+1}^nR_k
\label{pn0t1flow}\eeq
By iterating the relation (\ref{pratio}) using (\ref{Qndef}) it follows that
\beq
p_{n,0}\,l_{n,0}=(-1)^{n+1}\,(R_n-1)
\label{plRnrel}\eeq
where we have defined $R_0\equiv0$. Substituting (\ref{plRnrel}) into
(\ref{pn0t1flow}) we obtain
\beq
\frac{\partial p_{n,0}}{\partial
t_1}=p_{n+1,0}\,R_n=p_{n+1,0}\,\Bigl(1+(-1)^{n+1}\,p_{n,0}\,l_{n,0}\Bigr)
\label{pn0t1flowfin}\eeq
Similarly, one can derive the evolution equation
\beq
\frac{\partial l_{n,0}}{\partial
t_1}=-l_{n-1,0}\,\Bigl(1+(-1)^{n+1}\,p_{n,0}\,l_{n,0}\Bigr)
\label{ln0t1flow}\eeq

The flow equations (\ref{pn0t1flowfin}) and (\ref{ln0t1flow}) are the first
members of an integrable Hamiltonian system known as the Toeplitz chain
hierarchy \cite{adler}. The general evolution equations may be similarly
derived using the flow equations of the previous subsection and are given by
the Hamilton equations of motion
\beq
\frac{\partial p_{n,0}}{\partial
t_k}=\Bigl(1+(-1)^{n+1}\,p_{n,0}\,l_{n,0}\Bigr)\,\frac{\partial H_k}{\partial
l_{n,0}}~~~~~~,~~~~~~\frac{\partial l_{n,0}}{\partial
t_k}=-\Bigl(1+(-1)^{n+1}\,p_{n,0}\,l_{n,0}\Bigr)\,
\frac{\partial H_k}{\partial p_{n,0}}
\label{Toeplitzchain}\eeq
for the system of Hamiltonians
\beq
H_k=-\frac1k\,\Tr_{(N)}\left(\Q^k\right)=-\frac1k\,\frac{\partial\log
Z_N}{\partial t_k}~~~~~~,~~~~~~k\geq1
\label{Hams}\eeq
and the symplectic structure
\beq
\omega=\sum_{n\geq1}\frac{dp_{n,0}\wedge
dl_{n,0}}{1+(-1)^{n+1}\,p_{n,0}\,l_{n,0}}
\label{omegasympl}\eeq
The Toeplitz chain is a particular reduction of the {\it two-dimensional} Toda
lattice hierarchy, the latter of which characterizes generic matrix integrals
\cite{aaaaa,inthi}. Thus the integrable hierarchy to which the adjoint fermion
one-matrix model belongs is not the reduction of a one-dimensional (Toda)
hierarchy, but rather of a two-dimensional one. This is of course anticipated
from the lattice interpretation of the fermionic matrix model and its
associated doubling of degrees of freedom due to the Penner matter states.

However, the Toeplitz chain hierarchy can be viewed as a deformation of the
standard Toda chain hierarchy. To see this, we use (\ref{diffeq2x}) for $k=1$
and (\ref{calPQR}) for $m=n$ to obtain the flow equation
\beq
\frac{\partial\log h_n}{\partial t_1}=[\Q]_{nn}=Q_n\Bigl(R_{n+1}-1\Bigr)
\label{hnflowt1}\eeq
On the other hand, by using (\ref{diffeq1x}) for $k=1$ and (\ref{calPQR}) for
$m=n-1$ we have
\beq
\frac{\partial[\Q]_{nn}}{\partial t_1}=Q_n\Bigl[Q_{n+1}R_{n+1}(R_{n+2}-1)
-Q_{n-1}R_n(R_{n+1}-1)\Bigr]
\label{Qnnflowt1}\eeq
The two flow equations (\ref{hnflowt1}) and (\ref{Qnnflowt1}) may be combined
together to give
\beq
\frac{\partial^2\log h_n}{\partial t_1^2}=\frac{\partial\log h_n}{\partial
t_1}\left(\frac{R_{n+1}}{R_{n+1}-1}\,\frac{\partial\log h_{n+1}}{\partial
t_1}-\frac{R_n}{R_n-1}\,\frac{\partial\log h_{n-1}}{\partial t_1}\right)
\label{hnflowt1order2}\eeq
By introducing time-dependent functions $q_n[\,\vec t~]$ through
\beq
h_n[\,\vec t~]=(-1)^n\,\varepsilon^{2n}~\e^{t_1/\varepsilon}~\e^{q_n[\,\vec
t~]}
\label{qndef}\eeq
where $\varepsilon$ is a time-independent parameter, we can write
(\ref{hnflowt1order2}) as
\bea
\frac{\partial^2q_n}{\partial t_1^2}&=&\left(1+\varepsilon\,\frac{\partial
q_n}{\partial t_1}\right)\left(1+\varepsilon\,\frac{\partial q_{n+1}}{\partial
t_1}\right)\,\frac{\e^{q_{n+1}-q_n}}{1+\varepsilon^2~\e^{q_{n+1}-q_n}}\nn\\&
&-\,\left(1+\varepsilon\,\frac{\partial q_{n-1}}{\partial
t_1}\right)\left(1+\varepsilon\,\frac{\partial q_n}{\partial
t_1}\right)\,\frac{\e^{q_n-q_{n-1}}}{1+\varepsilon^2~\e^{q_n-q_{n-1}}}
\label{RTCeqn}\eea
The differential equation (\ref{RTCeqn}) is the first member of an integrable
hierarchy known as the {\it relativistic} Toda chain \cite{relToda}. In the
``non-relativistic'' limit $\varepsilon\to0$, it reduces to the Toda chain
hierarchy which characterizes generic Hermitian one-matrix models. Thus, the
partition function of the adjoint fermion one-matrix model is a $\tau$-function
of not the Toda chain hierarchy, but rather of its deformation to the
relativistic Toda chain. The latter chain is itself a reduction of the
two-dimensional Toda lattice. From this point of view, the fermionic matrix
model may be thought of as a ``deformation'' of the Hermitian one-matrix model,
the role of the deformation being played by the Penner interaction.

The remaining differential equations of the relativistic Toda chain hierarchy
are given by the Lax equations (\ref{Qktimederivs}), by (\ref{diffeq2x}), and
by exploiting the Jacobi properties (\ref{JacobiQ}) and (\ref{QE}) of the
operator $\Q$. It is important to realize that, because the present model
contains only a single set $\vec t$ of times, the ordinary Toda lattice does
not itself appear in the hierarchy satisfied by the partition function. Thus
the standard Toda lattice structure which underlies all integrable models (and
in particular matrix integrals) only appears in a very subtle way through the
reductions
obtained above by eliminating one set of its time variables. This reduced
structure leads to a certain degeneracy in the $\tau$-function as compared to
the usual two-dimensional Toda lattice structure. Of course, the relations to
integrable hierarchies described in this subsection are only `kinematical', as
they merely rely on the very basic recursion properties satisfied by the
orthogonal polynomials. To incorporate the `dynamical' aspects, and in
particular the effects of the Penner potential, we need to impose the
constraints implied by the string equations. Indeed, there are many
$\tau$-function solutions of the above hierarchies, and the string equations
select the one appropriate to describe the nonperturbative dynamics of the
adjoint fermion one-matrix model. This is the topic of the next subsection.

\subsection{Virasoro Constraints}

We will now begin examining the constraints imposed on the fermionic
$\tau$-function
$Z_N[\,\vec t~]$ as dictated by the string equations. Proceeding as in
(\ref{eqmotionderiv}) for the operators $\P\,\Q^{n+1}$, $n\geq-1$, we can
express the string equation (\ref{opid1}) as an infinite system of equations
\beq
\Tr_{(N)}\left[\left(\P+\L^\dagger\,\Q^{-2}-(N+1)\,\Q^{-1}+\sum_{k\geq1}
kt_k\,\Q^{k-1}\right)\Q^{n+1}\right]=0~~~~~~,~~~~~~n\geq-1
\label{TrNn-1}\eeq
The strategy now is to rewrite the equations of motion (\ref{TrNn-1}) using the
flow equations (\ref{ZNflow}) and (\ref{ZNflow2}). In this way, the string
equations will be represented as the annihilation of the partition function by
a system of differential operators in the coupling constants of the
potential $V$. In this subsection we will deal with the cases $n\geq0$.

Let us start with the $n=0$ string equation of (\ref{TrNn-1}). Using the
canonical commutation relations (\ref{cancommrel}) and the matrix elements
\beq
h_n[\Q\,\P]_{nn}=\Bigl\la z\,\Phi_n'(z)\Bigm|\Lambda_n(z)\Bigr\ra=nh_n
\label{QPnn}\eeq
we may compute the first trace in (\ref{TrNn-1}) for $n=0$ as
\beq
\Tr_{(N)}(\P\,\Q)=N+\sum_{n=0}^{N-1}n=\frac{N(N+1)}2
\label{TrPQ}\eeq
The second trace may be similarly computed by using
\beq
h_n\left[\L^\dagger\,\Q^{-1}\right]_{nn}=\Bigl\la\Phi_n(z)\Bigm|z
\,\Lambda_n'(z)\Bigr\ra=-nh_n
\label{barPQinvnn}\eeq
to get
\beq
\Tr_{(N)}\left(\L^\dagger\,\Q^{-1}\right)=-\frac{N(N-1)}2
\label{TrbarPQinv}\eeq
By substituting (\ref{TrPQ}) and (\ref{TrbarPQinv}), and using (\ref{ZNflow}),
we can represent the $n=0$ constraint of the system (\ref{TrNn-1}) as the flow
equation
\beq
\left(\sum_{k\geq1}kt_k\,\frac\partial{\partial t_k}-N^2\right)Z_N[\,\vec t~]=0
\label{Vir0}\eeq

Next, let us consider the $n=1$ constraint of (\ref{TrNn-1}). For the first
trace, we use (\ref{norm}) and (\ref{monicexp}) to compute the matrix elements
\bea
h_n\left[\Q^2\,\P\right]_{nn}&=&\Bigl\la z^2\,\Phi_n'(z)\Bigm|\Lambda_n(z)
\Bigr\ra\nn\\&=&\Bigl\la n\,z^{n+1}-(n-1)p_{n,n-1}\,z^n+O\left(z^{n-1}
\right)\Bigm|
\Lambda_n(z)\Bigr\ra\nn\\&=&\Bigl[np_{n+1,n}-(n-1)p_{n,n-1}\Bigr]h_n
\label{Q2Pnn}\eea
Using (\ref{pnn-1sol}) and the canonical commutator (\ref{cancommrel}) we
thereby arrive after a little algebra at
\beq
\Tr_{(N)}\left(\P\,\Q^2\right)=2\,\Tr_{(N)}(\Q)+\sum_{n=1}^{N-1}
\Bigl(n[\Q]_{nn}+\Tr_{(n)}(\Q)\Bigr)=(N+1)\,\Tr_{(N)}(\Q)
\label{TrQ2P}\eeq
Since $\left[\L^\dagger\right]_{nn}=0$, upon substituting (\ref{TrQ2P}) into
(\ref{TrNn-1}) we find that the terms linear in $\Q$ cancel out and applying
the flow equation (\ref{ZNflow}) we see that the $n=1$ constraint can be
written as
\beq
\sum_{k\geq1}kt_k\,\frac\partial{\partial t_{k+1}}~Z_N[\,\vec t~]=0
\label{Vir1}\eeq

Now we move on to the $n=2$ string equation of (\ref{TrNn-1}).
Again using (\ref{norm}) and (\ref{monicexp}) we may compute
\bea
h_n\left[\Q^3\,\P\right]_{nn}&=&\Bigl\la z^3\,\Phi_n'(z)\Bigm|\Lambda_n(z)
\Bigr\ra\nn\\&=&\Bigl\la n\,z^{n+2}-(n-1)p_{n,n-1}\,z^{n+1}-(n-2)p_{n,n-2}\,z^n
+O\left(z^{n-1}\right)
\Bigm|\Lambda_n(z)\Bigr\ra\nn\\&=&\Bigl[n(p_{n+2,n}+p_{n+2,n+1}
p_{n+1,n})-(n-2)p_{n,n-2}-(n-1)p_{n+1,n}p_{n,n-1}\Bigr]h_n\nn\\& &
\label{Q3Pnn}\eea
Using (\ref{pnn-1sol}) we can iterate the relation (\ref{pnkit}) for $k=n-2$ to
get
\beq
p_{n,n-2}=\sum_{k=1}^{n-1}\Bigl([\Q]_{k,k-1}+[\Q]_{kk}\,\Tr_{(k)}(\Q)\Bigr)
\label{pnn-2sol}\eeq
Substituting (\ref{pnn-2sol}) and (\ref{pnn-1sol}) into (\ref{Q3Pnn}) and using
(\ref{cancommrel}), we arrive after some algebra at
\bea
\Tr_{(N)}\left(\P\,\Q^3\right)&=&3\,\Tr_{(N)}\left(\Q^2\right)+(N-1)[\Q]
_{N,N-1}+(N+1)\Bigl(\Tr_{(N)}(\Q)\Bigr)^2\nn\\& &+\,(2N-3)\sum_{n=1}^{N-1}
\Bigl([\Q]_{n,n-1}-[\Q]_{nn}\,\Tr_{(n)}(\Q)\Bigr)
\label{TrPQ3}\eea
The sums in (\ref{TrPQ3}) can be simplified by using the Jacobi property
(\ref{JacobiQ}) of the operator $\Q$ to get
\beq
\Tr_{(N)}\Bigl(\Q^2\Bigr)=[\Q]_{N,N-1}+2\sum_{n=1}^{N-1}[\Q]_{n,n-1}
+\sum_{n=0}^{N-1}\Bigl([\Q]_{nn}\Bigr)^2
\label{TrQ2}\eeq
and in this way we arrive at an expression for the first trace in the $n=2$
equation of (\ref{TrNn-1}),
\beq
\Tr_{(N)}\left(\P\,\Q^3\right)=\left(N+\mbox{$\frac32$}\right)\,\Tr_{(N)}
\left(\Q^2\right)
+\mbox{$\frac12$}\,[\Q]_{N,N-1}+\mbox{$\frac12$}\,\Bigl(\Tr_{(N)}(\Q)\Bigr)^2
\label{TrPQ3fin}\eeq

For the second trace in (\ref{TrNn-1}), we use (\ref{norm}) to compute the
matrix elements
\bea
h_n\left[\L^\dagger\,\Q\right]_{nn}&=&\Bigl\la\Phi_n(z)\Bigm|z^3\,\Lambda_n'(z)
\Bigr\ra\nn\\&=&\Bigl\la\Phi_n(z)\Bigm|-n\,z^{-(n-2)}+O\left(z^{-(n-3)}
\right)+l_{n,1}\,z
\Bigr\ra\nn\\&=&l_{n,1}\,\Bigl\la z\,\Phi_n(z)\Bigm|1\Bigr\ra
\label{barPQnn}\eea
Using (\ref{zPhiexpl}) we obtain
\beq
\left[\L^\dagger\,\Q\right]_{nn}=S_nQ_n(1-R_{n+1})
\label{barPQSn}\eeq
where we have defined
\beq
S_n=\frac{l_{n,1}}{l_{n,0}}
\label{Sn}\eeq
By equating the coefficients of the $z^{-1}$ terms on both sides of the
three-term recursion relation (\ref{3termLambda}) for the $\Lambda$
polynomials, we can write down an iterative equation for the coefficients
(\ref{Sn}),
\beq
S_n=R_nQ_{n-1}+S_{n+1}-Q_n
\label{Snit}\eeq
which has solution
\beq
S_n=Q_{n-1}R_n+\sum_{k=0}^{n-1}Q_k(1-R_{k+1})
\label{Snsol}\eeq
Substituting (\ref{Snsol}) into (\ref{barPQSn}) and using (\ref{calPQR}) for
$m=n-1$ and $m=n$, we therefore find
\beq
\left[\L^\dagger\,\Q\right]_{nn}=[\Q]_{nn}\,\Tr_{(n)}(\Q)-[\Q]_{n,n-1}
\label{barPQnnfin}\eeq
Using the identity (\ref{TrQ2}) we then arrive at an
expression for the second trace in the $n=2$ equation of (\ref{TrNn-1}),
\beq
\Tr_{(N)}\left(\L^\dagger\,\Q\right)=\mbox{$\frac12$}\,[\Q]_{N,N-1}+\mbox
{$\frac12$}\,\Bigl(\Tr_{(N)}(\Q)\Bigr)^2-\mbox{$\frac12$}\,
\Tr_{(N)}\left(\Q^2\right)
\label{TrbarPQ}\eeq

Upon substitution of (\ref{TrPQ3fin}) and (\ref{TrbarPQ}) into (\ref{TrNn-1})
for $n=2$, we see that the terms quadratic in $\Q$ cancel out. The remaining
terms can be simplified using the flow equations (\ref{ZNflow}) and
(\ref{ZNflow2}), which leads to the $n=2$ constraint equation
\beq
\left(\sum_{k\geq1}kt_k\,\frac\partial{\partial t_{k+2}}+\frac{\partial^2}
{\partial t_1^2}\right)Z_N[\,\vec t~]=0
\label{Vir2}\eeq
This procedure can be easily generalized to higher orders $n\geq3$. However, it
is possible to determine the general relation from the first three equations
(\ref{Vir0}), (\ref{Vir1}) and (\ref{Vir2}) by using the fact that the
commutator bracket of the pertinent differential operators must also annihilate
the partition function and thereby using commutators of the above operators to
generate the higher order ones. The final result can be expressed in a more
familiar form by introducing the zeroth time $t_0$ through the flow equation
\beq
\frac{\partial Z_N}{\partial t_0}=NZ_N
\label{t0def}\eeq
for the partition function.

In this way we find that the string equations (\ref{TrNn-1}) for $n\geq0$ can
be written as the set of discrete Virasoro constraints
\beq
L_n\,Z_N[\,\vec t~]=0~~~~~~,~~~~~~n\geq0
\label{Virconstr}\eeq
where the second order linear differential operators
\beq
L_n=\sum_{k\geq1}kt_k\,\frac\partial{\partial t_{k+n}}+\sum_{k=0}^n
\frac{\partial}{\partial t_k}\,\frac\partial
{\partial t_{n-k}}-2N\,\frac\partial{\partial t_n}
\label{Virops}\eeq
satisfy the Witt algebra
\beq
\Bigl[L_n\,,\,L_m\Bigr]=(n-m)\,L_{n+m}
\label{Wittalg}\eeq
for $n,m\in\IZ^+$. These Virasoro constraints are the usual ones for the
fermionic matrix model \cite{mz,ss1} and they are identical to those of the
Hermitian Penner matrix model \cite{pennerloop} as defined in (\ref{ZNH}) 
with $\alpha=-2$. They represent the full system
of equations of motion of the matrix model and can be equivalently derived by
demanding the invariance of the matrix integral under arbitrary changes of the
matrix variables \cite{mz,ss1}. The first set
of operators in (\ref{Virops}) comes from the variation of the potential $V$ in
the action, while the second one comes from the Jacobian of the change of
matrix integration measure. The first two terms in (\ref{Virops}) therefore
coincide with the standard Virasoro generators of generic Hermitian one-matrix
models \cite{aaaaa}. The last operator in (\ref{Virops}) comes from the
variation of the logarithmic Penner interaction.

It is a remarkable fact that the loop equations of the adjoint fermion
one-matrix model (\ref{1mat}), the Hermitian Penner one-matrix model 
((\ref{ZNH}) with $\alpha =-2$), and the unitary Penner
matrix model (\ref{ummdef1})
are all equivalent. However, the equivalence between the Hermitian
model and the other two matrix models only holds order by order in the
$\frac1N$ expansions of the matrix integrals. Although the models are
equivalent at $N=\infty$, beyond leading order in the large $N$ expansion the
loop equations should be solved with different boundary conditions and the
solutions are different. The fermionic and unitary equations should be solved
with boundary conditions appropriate to a perturbative Gaussian limit, in order
that the models admit interpretations in terms of random surfaces, while the
Hermitian ones should be solved with boundary conditions appropriate to recover
the Penner model of the discretized moduli space of Riemann surfaces
\cite{acm,penner}. The distinction of the fermionic and unitary matrix
integrals from the Hermitian one is evident from the structure of the finite
$N$ correlators (see appendix A). In fact, it is only in the large $N$ limit
that a full Virasoro symmetry is realized in these models, as well as an
infinite hierarchy of generalized KP differential equations. This makes the
large $N$ limit a very natural ingredient of the fermionic matrix model, as is
also apparent from its fat-graph interpretation of section~2.2.

\subsection{Origin of the Painlev\'e Expansion}

In the previous subsection we have shown that the Schwinger-Dyson equations of
the fermionic matrix model are generated by the positive Borel subalgebra of a
Virasoro algebra of vanishing central charge. There is a nice algebraic way to
characterize the effect of the Penner interaction potential in the Virasoro
generators (\ref{Virops}). Let us write them as $L_n=L_n^{\rm H}+T_n$, where
$L_n^{\rm H}$
are the standard Virasoro generators of a Hermitian one-matrix model with
potential $V$, and
\beq
T_n=-2N\,\frac\partial{\partial t_n}
\label{Tndef}\eeq
are the generators of translations $t_n\mapsto t_n-2N\varepsilon$ in coupling
constant space. Together these operators  generate
the symmetry algebra
\bea
\Bigl[T_n\,,\,T_m\Bigr]&=&0\nn\\\Bigl[L_n^{\rm H}\,,\,L_m^{\rm H}\Bigr]
&=&(n-m)\,L_{n+m}^{\rm H}\nn\\\Bigl[T_n\,,\,L_m^{\rm H}\Bigr]&=&
-2Nn\,T_{n+m}
\label{mastersym}\eea
The commutation relations (\ref{mastersym}) characterize a master symmetry of
the generalized KdV hierarchy \cite{tau}. We see then that the Virasoro
constraints of the fermionic matrix model constitute a deformation of those for
Hermitian one-matrix models by the master symmetry algebra generators of the
KdV flow equations. Note that the particular symmetry generated by the
operators (\ref{Tndef}) is restricted to translations in units of a discrete
``lattice spacing'' $-2N$, indicating what sort of reduction of this master
symmetry is taken. These Virasoro constraints describe the desired reduction of
the fermionic $\tau$-function, and they lead immediately to the desired
geometrical interpretation of the Painlev\'e expansion of the adjoint fermion
one-matrix model. The Painlev\'e equations may be characterized as following
from a reduction of the $\tau$-function of the integrable Toeplitz (or
relativistic Toda) chain hierarchy, subject to the Virasoro symmetry obtained
in the previous subsection. They yield a stability condition $L_n\,\tau_N=0$ on
the points of the model Grassmannian, which are associated with the
Baker-Akhiezer functions (\ref{BAdef}), that select a particular class of
transcendental solutions to the KP equations.

It is a standard fact that a set of operators of the form (\ref{Virops}) can be
embedded into a Virasoro algebra of non-vanishing central
charge. However, an important aspect of the present approach to the fermionic
matrix model is that there are in fact extra dynamical constraints imposed on
the partition function which can be attributed to the invertibility of the Lax
operator $\Q$. For instance, the constraint (\ref{TrNn-1}) makes perfect sense
for $n=-1$, and the calculation of the trace of the operator
$\L^\dagger\,\Q^{-2}$ proceeds in an analogous manner to that of (\ref{Q2Pnn})
with the result
\beq
\Tr_{(N)}\left(\L^\dagger\,\Q^{-2}\right)=-(N-1)\,\Tr_{(N)}\left(\Q^{-1}\right)
\label{barPQ2nn}\eeq
Since $[\P]_{nn}=0$, we see that on substitution of (\ref{barPQ2nn}) into
(\ref{TrNn-1}) for $n=-1$ we produce an extra non-vanishing term
$-2N\,\Tr_{(N)}(\Q^{-1})$. This trace can be written in an operator form as
follows. From (\ref{calLexpl}) with $m=n$ and (\ref{hnflowt1}) we have
\beq
\left[\Q^{-1}\right]_{nn}=\Bigl(R_n-1\Bigr)\Bigl(R_{n+1}-1\Bigr)\left(
\frac{\partial\log h_n}{\partial t_1}\right)^{-1}
\label{Qinvnn}\eeq
This quantity is independent of the first time $t_1$, since on using the flow
equation (\ref{Mflow}) for $k=1$ and the Jacobi properties of the operators
$\Q$ and $\M$ we find $\partial[\Q^{-1}]_{nn}/\partial t_1=0$. This suggests
defining a negative time $t_{-1}$ such that $\Tr_{(N)}(\Q^{-1})$ is represented
as an operator acting on the partition function as
\bea
\frac{DZ_N}{Dt_{-1}}&=&\sum_{n=1}^{N-1}\frac1{\tau_n\tau_{n+1}}
\,\frac{\Bigl(\tau_{n+1}\tau_{n-1}-\tau_n^2\Bigr)
\left(\tau_{n+2}\tau_n-\tau_{n+1}^2\right)}
{\tau_n\,\frac{\partial\tau_{n+1}}{\partial t_1}-\tau_{n+1}\,
\frac{\partial\tau_n}{\partial t_1}}\nn\\\frac\partial{\partial t_1}\,
\frac{DZ_N}{Dt_{-1}}&=&0
\label{t-1flowdef}\eea
where we have used (\ref{taundef}). Then the constraint (\ref{TrNn-1}) for
$n=-1$ can be written as
\beq
\left(\sum_{k\geq2}kt_k\,\frac\partial{\partial t_{k-1}}+Nt_1-2N\,\frac D
{Dt_{-1}}\right)Z_N[\,\vec t~]=0
\label{Vir-1}\eeq

In fact, there are infinitely many negatively moded constraints such as this,
associated with higher powers $\Q^{-m}$, $m\geq1$. However, the operators such
as that in (\ref{Vir-1}) which are thereby generated do not form a closed
algebra among themselves or with the operators (\ref{Virops}). Indeed, this set
of
constraints arises from the previously mentioned reduction from a
two-dimensional lattice (containing both negative and positive time parameters)
to a (deformed) chain by eliminating all negative couplings. As usual, such
reductions break the full Virasoro symmetry of the original integrable model
and only the positive Borel subalgebra survives here as a symmetry of the
matrix model. The negatively moded constraints are therefore merely an artifact
of the lattice reduction or equivalently the chain deformation. In practical 
terms, all observables in the fermionic model are described by observables
conjugate to positive times in the unitary matrix model.  Observables conjugate
to negative times exist but do not appear to play a role in the equivalence.

We can understand all of these matters more clearly by rewriting the Virasoro
constraints (\ref{Virconstr}) in terms of the Baker-Akhiezer functions
(\ref{BAdef}). For this, we note first of all that (\ref{Phiflow}) and
(\ref{diffeq2x}) imply that they obey the flow equations
\beq
\frac{\partial\Psi_n[\,\vec t\,;z]}{\partial t_k}=\frac12\left(\Q_+^k-\Q_-^k-
\left[\Q^k\right]_{nn}\right)\,\Psi_n[\,\vec t\,;z]
\label{BAflow}\eeq
Now let us compute the action of the scale transformation generator
$z\,\frac\partial{\partial z}$ in eigenvalue space on the Baker-Akhiezer
functions. We have
\beq
z\,\frac{\partial\Psi_n[\,\vec t\,;z]}{\partial z}=\left(-\frac{N+3}2+\frac
N2\,z\,V'(z)\right)\Psi_n[\,\vec t\,;z]+\frac{\e^{\frac N2\,V(z)}}
{z^{\frac{N+1}2}}\,\P\,\Q\,\Phi_n(z)
\label{scaleaction}\eeq
To evaluate the last term in (\ref{scaleaction}), we have to take into account
the triangularity of the operator $\P\,\Q$. From $\P=\P_-$ and the Jacobi
property (\ref{JacobiQ}), we have
\beq
\P\,\Q=(\P\,\Q)_-+\sum_{n\geq0}[\P\,\Q]_{nn}\,{\bf E}_{n,n}
\label{PQ-diag}\eeq
We can therefore evaluate the action of $\P\,\Q$ on $\Phi_n(z)$ by using the
string equation (\ref{opid1}) multiplied on the right by $\Q$, but eliminating
the pure upper triangular part of both sides of the equation. Using
(\ref{barPQinvnn}) and the fact that
\beq
\left(\L^\dagger\,\Q^{-1}\right)_-=0
\label{barPQ-0}\eeq
which can be derived similarly to the other matrix elements of this section, we
can write (\ref{scaleaction}) as
\bea
z\,\frac{\partial\Psi_n[\,\vec t\,;z]}{\partial z}&=&
\frac12\,\Bigl(2n+N-1\Bigr)\Psi_n[\,\vec t\,;z]\nn\\& &
+\,\frac N2\left[\Bigl(\Q\,V'(\Q)\Bigr)_+-\Bigl(\Q\,V'(\Q)
\Bigr)_--\Bigl[\Q\,V'(\Q)\Bigr]_{nn}\right]\Psi_n[\,\vec t\,;z]
\label{scalesimpl}\eea
The last term in (\ref{scalesimpl}) can be rewritten using the flow equation
(\ref{BAflow}), leading to the eigenvalue equation
\beq
L_0[\,\vec t\,;z]\,\Psi_n[\,\vec t\,;z]=-n\,\Psi_n[\,\vec t\,;z]
\label{L0BA}\eeq
where we have defined the Virasoro operator
\beq
L_0[\,\vec t\,;z]=\sum_{k\geq1}kt_k\,\frac\partial{\partial t_k}-z\,\frac
\partial{\partial z}+\frac{N-1}2
\label{L0eigendef}\eeq

The eigenvalue equation (\ref{L0BA}) shows that the lowest Virasoro constraint
is indeed a true symmetry of the system, in that the operators $\Q$ and
$L_0[\,\vec t\,;z]$ are simultaneously diagonalizable in the basis of
bi-orthonormal Baker-Akhiezer functions. The same is true of all higher
Virasoro constraints. This shows precisely what sort of constraints the
auxiliary eigenvalue problem (\ref{eigeneqn}) must possess in order to
correctly reconstruct the fermionic $\tau$-function. However, an analogous
computation using the translation generator $\frac\partial{\partial z}$ on
eigenvalue space leads to the result
\beq
\left(\sum_{k\geq1}kt_k\,\frac\partial{\partial t_{k-1}}-\frac\partial
{\partial z}-\frac{N+1}{2z}\right)\Psi_n[\,\vec t\,;z]=-(n+N)R_n\,
\Psi_{n-1}[\,\vec t\,;z]
\label{L-1BA}\eeq
which shows that there is no corresponding $L_{-1}$ operator which commutes
with the Lax operator $\Q$. This means that there is no translational symmetry
in the system, and the leading equations of motion at large $N$ are determined
instead by the generator $L_0$ of scale transformations of the system. These
equations can be written in a form comparable to those of sections 3 and 4 by
writing a consistency condition between the eigenvalue equations
(\ref{eigeneqn}) and (\ref{L0BA}) in the form
\beq
\Bigl(L_0[\,\vec t\,;z]+n\Bigr)\Bigl(\Q-z\Bigr)\Psi_n[\,\vec t\,;z]=0
\label{eigenconscondn}\eeq
Expanding (\ref{eigenconscondn}) using (\ref{L0BA}), (\ref{L0eigendef}) and
completeness of the Baker-Akhiezer functions then leads to the system of
equations
\beq
(1+n-m)\,[\Q]_{nm}+\sum_{k\geq1}kt_k\,\frac{\partial[\Q]_{nm}}{\partial t_k}=0
\label{conscondnexp}\eeq
In particular, setting $n=m$ in (\ref{conscondnexp}) and using the flow
equation (\ref{diffeq1x}), we have
\beq
[\Q]_{nn}=\sum_{k\geq1}kt_k\left(\Bigl[\Q^k\Bigr]_{n,n-1}-\Bigl[\Q^k\Bigr]
_{n+1,n}\right)
\label{Qstringeqn}\eeq

What this means is that it is the operator $\P\,\Q$ (along with its conjugate
$\L^\dagger\,\Q^{-1}$) which plays the fundamental role in the dynamics of the
adjoint fermion one-matrix model. Thus it is the scale invariance of the system
which leads to the fundamental double scaling equations in the large $N$ limit.
The translational symmetry generated by the operator $\P$ itself (and its
conjugate $\L^\dagger\,\Q^{-2}$) is an extra constraint on the theory which
plays no immediate role in the continuum limit of the matrix model. By
considering the matrix elements
$\la\Phi_n(z)|z\,\bigl[z\,\Lambda_m(z)\bigr]'\ra$, it is possible to infer the
commutation relation $\left[\Q\,,\,\L^\dagger\,\Q^{-1}\right]=\Q$. Using the
canonical commutator (\ref{cancommrel}), it follows that the fundamental
symmetry operator of the system is given by
\beq
\Pi=\mbox{$\frac12$}\,\left(\P\,\Q-\L^\dagger\,\Q^{-1}\right)
\label{Pidef}\eeq
which together with the Lax operator $\Q$ obeys the non-canonical commutation
relation
\beq
\Bigl[\Pi\,,\,\Q\Bigr]=\Q
\label{PiQcomm}\eeq
The operator expression (\ref{PiQcomm}) defines the appropriate string
equations of the fermionic matrix model. The upper and lower triangular parts
of the scaling operator (\ref{Pidef}) may be computed by using (\ref{opid1}),
(\ref{QPnn}), (\ref{PQ-diag}) and (\ref{barPQ-0}) to get
\bea
\Pi_+&=&\frac12\,\sum_{n\geq0}(n-N)\,{\bf E}_{n,n}+\frac12\,
\sum_{k\geq1}kt_k\,\Q_+^k\nn\\\Pi_-&=&-\frac12\,\sum_{k\geq1}kt_k\,\Q_-^k
\label{Piexpl}\eea
Substituting (\ref{Piexpl}) into the string equation (\ref{PiQcomm}) and using
the KP equation (\ref{Qflow}), we arrive at (\ref{conscondnexp}).

It is the equation (\ref{Qstringeqn}) which produces the novel Painlev\'e
expansions of the fermionic matrix model. For instance, in the case of the
quadratic potential studied in section 4, it is implicit in the manipulations
carried out in section 4.2. In fact, the structural form of
(\ref{conscondnexp}) is identical to the so-called ``automodel'' constraints which arise in
the usual Hermitian one-matrix models \cite{aaaaa}. On the other hand, in the
continuum limit $\Q$ becomes a differential operator of a certain finite degree
and the string equations (\ref{PiQcomm}) can be translated into an equation for
pseudo-differential operators which satisfy a generalized KdV hierarchy of flow
equations \cite{fgz}. General solutions of string equations such as
(\ref{PiQcomm}) have been studied before, in the case that the double scaling
limit of $\Q$ is described by a Schr\"odinger operator, with stable, pole-free
solutions for the corresponding string susceptibility \cite{djm}. We see
therefore that the string equations of the fermionic one-matrix model lead to a
Painlev\'e differential equation which has the same structure as that in the
usual Hermitian one-matrix models, except that now the automodel form of the
equations produce an alternating genus expansion. Moreover, the structural form
(\ref{PiQcomm}) of the string equations allow us to restrict to positive values
of the double-scaling variable $\xi$ and thereby obtain pole-free solutions for
the free energy of the fermionic matrix model.

In the present case, the details of the continuum limit of the operator $\Q$
are somewhat involved. However, the discrete equations (\ref{conscondnexp}),
(\ref{Qstringeqn}) and (\ref{PiQcomm}) completely characterize the double
scaled equations for the partition function in the large $N$ limit and the
ensuing topological expansion. They serve as the starting point for a
characterization of the differential hierarchies of the fermionic one-matrix
model. From the generic structure of the flow equations obtained in this
section, we can expect that the continuum equations satisfied by the partition
function of ``fermionic quantum gravity'' are closely related to the usual KdV
flow structure of two-dimensional quantum gravity in terms of Gelfand-Dikii
differential polynomials \cite{fgz,2Dbook}.

\subsection*{Acknowledgments}

We thank J. Ambj\o rn, C. Kristjansen, Y. Makeenko, G. Semenoff and J. Wheater
for helpful discussions. We also thank A.A. Kapaev, M. Matone and P. Zinn-Justin for comments
on the manuscript. The work of L.D.P. was supported in part by the
Natural Sciences and Engineering Research Council of Canada and NSF grant
PHY98-02484. The work of R.J.S. was supported in part by the Danish Natural
Science Research Council.

\setcounter{section}{0}

\appendix{Spectral Correlation Functions}

In this appendix we will briefly describe some properties of correlators of the
fermionic one-matrix model within the formalism of this paper. The symmetries
of the matrix integral (\ref{1mat}) restrict its observables to those which are
invariant functions of $\ps2$ \cite{ss1}. Connected correlation functions
may be generated by taking derivatives of the free energy with respect to the
coupling constants of the potential $V$,
\beq
\left\langle\prod_{j=1}^L\tr\left(\ps2\right)^{p_j}\right\rangle_{\rm F\,,\,
conn}=\frac1{N^L}\,\prod_{j=1}^Lp_j\,\frac\partial{\partial g_{p_j}}~\log Z_N
\label{Fconndef}\eeq
where the normalized fermionic correlation functions are defined by
\beq
\left\langle f\left(\ps2\right)\right\rangle_{\rm F}\equiv
\frac1{Z_N}\,\int\limits_{{\rm Gr}(N)^c}d\psi~d\bar\psi~f\left(\ps2\right)
{}~\e^{N\,\tr\,V(\ps2)}
\label{Fcorrfndef}\eeq
and in (\ref{Fconndef}) it is understood that, if necessary, a set of
auxiliary coupling constants $g_k$ are introduced into the potential $V$ and
then set to zero after differentiation. Given the equivalence of the generating
function $\log Z_N$ for the connected correlators of the fermionic and unitary
matrix models, we therefore also have complete equivalence of their
observables,
\beq
\left\langle\prod_{j=1}^L\tr\left(\ps2\right)^{p_j}\right\rangle_{\rm F\,,\,
conn}=\left\langle\prod_{j=1}^L\tr\,U^{p_j}\right\rangle_{\rm U\,,\,conn}
\label{FUconneq}\eeq
where the normalized unitary correlation functions are defined by
\beq
\Bigl\langle f(U)\Bigr\rangle_{\rm U}\equiv\frac{k_N}{Z_N}\,\int\limits_{U(N)}
[dU]~f(U)~\e^{N\,\tr\bigl(V(U)-\log U\bigr)}
\label{Ucorrfndef}\eeq

The correlation functions in (\ref{FUconneq}) may be evaluated as in
(\ref{ZNU1SUN})--(\ref{ZNrestr}). For example, one may readily compute
\beq
\left\la\tr\left(\ps2\right)^p\right\rangle_{\rm F}=\Bigl\langle\e^{ip\theta}~
\tr\,U_0^p\Bigr\rangle_{\rm U}=\frac{k_N}
{NZ_N}~\left[~\int\limits_{SU(N)}[dU_0]~
\tr\,U_0^p~\e^{N\,\tr\,V(U_0)}\,\right]_{C_1=N^2-p}
\label{trps2prep}\eeq
The restriction in (\ref{trps2prep}) of the character expansion of
$\e^{N\,\tr\,V(U_0)}$ to Young tableaux with $N^2-p$ filled boxes implies that
single trace correlators are non-vanishing only for $p\leq N^2$. The
finiteness of the set of non-vanishing correlators is of course natural as a
consequence of the anticommuting property of the matrices $\psi$ and
$\bar\psi$, but it is a dramatic result in the unitary matrix model.
Thus not only does the Penner interaction term in the
unitary matrix model (\ref{unitaryZN}) capture the finite nature of the
perturbation expansion of $Z_N$, but it also reduces the correlation functions
appropriately to reproduce the correct properties of the fermionic correlators
at finite $N$. Note that, geometrically, the operator $\tr(\ps2)^p$ inserts, on
the dual triangulated surface, a hole with $p$ boundary lengths. The lattice
expansion of such operators thereby generates fermionic ribbon graphs which are
dual to tesselations of Riemann surfaces with a given number of boundaries.
Again there are only finitely many such fat-graphs, because the maximum number
of boundaries that a given discretization can have is $N^2$.

The complete set of observables of the fermionic matrix model can be generated
by the joint probability distributions
\beq
\rho_n(z_1,\dots,z_n)=\frac{(N-n)!}{N!}\,\left\langle\prod_{k=1}^n\tr\,
\delta\left(z_k-\ps2\right)\right\rangle_{\rm F}
\label{jointprob}\eeq
where $z_k$ are points in the complex plane and $1\leq n\leq N$. When the
correlation function in (\ref{jointprob}) is mapped to the unitary matrix model
as described above, the points $z_k$ can be interpreted as eigenvalues of
unitary matrices. By diagonalizing the unitary matrices in (\ref{Ucorrfndef}),
using the identities
\beq
\Delta(z_1,\dots,z_N)=\det_{i,j}\Bigl[\Phi_{j-1}(z_i)\Bigr]~~~~~~,~~~~~~
\overline{\Delta(z_1,\dots,z_N)}=\det_{i,j}\Bigl[\Lambda_{j-1}(z_i)\Bigr]
\label{Deltaorthorels}\eeq
which follow from (\ref{vandermonde}) and (\ref{norm}), and by using
(\ref{rec}), it is straightforward to derive the determinant representation
\beq
\rho_n(z_1,\dots,z_n)=\frac{(N-n)!}{N!}\,\left\langle\prod_{k=1}^n\tr\,
\delta(z_k-U)\right\rangle_{\rm U}=\frac{(N-n)!}{N!}~
\det_{i,j}\Bigl[{\cal K}(z_i,z_j)\Bigr]
\label{rhondetcalK}\eeq
where ${\cal K}(z,z')$ is the spectral kernel which is defined in terms of the
orthogonal polynomials as
\beq
{\cal K}(z,z')=\frac{\e^{\frac N2\,\bigl(V(z)+V(z')\bigr)}}
{(z\,z')^{\frac{N+1}2}}
\,\sum_{n=0}^{N-1}\frac{\Phi_n(z)\,\Lambda_n(z')}{h_n}
\label{speckernel}\eeq

We see therefore that the problem of evaluating correlation functions of the
fermionic matrix model reduces to that of determining the spectral kernel
(\ref{speckernel}). It is possible to express it in a much simpler form by
deriving the appropriate generalization of the Christoffel-Darboux formula
\cite{mehta}. For this, we first need to derive a ``mixed'' recursion relation
between the $\Phi$ and $\Lambda$ polynomials. Consider the bi-orthogonality
relations
\bea
\Bigl\langle\Phi_{n+1}(z)-z\,\Phi_n(z)\Bigm|z^{-m}\Bigr\ra&=&
\Bigl\la\Phi_{n+1}(z)\Bigm|z^{-m}\Bigr\ra-\Bigl\la\Phi_n(z)\Bigm|z^{-(m-1)}
\Bigr\ra~=~0\nn\\\Bigl\la z^n\,\Lambda_n(z)\Bigm|z^{-m}\Bigr\ra&=&
\Bigl\la z^{n-m}\Bigm|\Lambda_n(z)\Bigr\ra~=~0
\label{mixedderiv}\eea
which are valid for $1\leq m\leq n$. From (\ref{mixedderiv}) it follows that
the two polynomials $\Phi_{n+1}(z)-z\,\Phi_n(z)$ and $z^n\,\Lambda_n(z)$ of
degree $n$ are equal up to some constant $c_n$. Equating the constant terms of
these polynomials gives $c_n=p_{n+1,0}$, and we arrive at the mixed three-term
recursion relation
\beq
\Phi_{n+1}(z)=z\,\Phi_n(z)+p_{n+1,0}\,z^n\,\Lambda_n(z)
\label{mixedrecrel1}\eeq
In an analogous way, it is possible to derive the recursion relation
\beq
\Lambda_{n+1}(z)=\frac1z\,\Lambda_n(z)+l_{n+1,0}\,\frac1{z^n}\,\Phi_n(z)
\label{mixedrecrel2}\eeq

We now multiply (\ref{mixedrecrel1}) through by $\Lambda_n(z')/h_n$ and
(\ref{mixedrecrel2}) with the index shift $n\to n-1$ and $z\to z'$ through by
$z'\,\Phi_n(z)/h_n$, subtract the resulting two equations, and then sum over
$n=1,\dots,N-1$. This yields
\bea
(z-z')\,\sum_{n=1}^{N-1}\frac{\Phi_n(z)\,\Lambda_n(z')}{h_n}&=&
\frac{\Phi_N(z)\,\Lambda_{N-1}(z')}{h_{N-1}}-\frac{\Phi_1(z)}{h_1}\nn\\& &-\,
\sum_{n=1}^{N-1}\frac1{h_k}\,\left[p_{n+1,0}\,z^n\,\Lambda_n(z)\,
\Lambda_n(z')-\frac{l_{n,0}}{z'^{n-1}}\,\Phi_n(z)\,\Phi_{n-1}(z')\right]
\nn\\& &
\label{mixeddiff}\eea
We can iterate the recursion equations (\ref{mixedrecrel1}) and
(\ref{mixedrecrel2}) to get
\beq
z^k\,\Lambda_k(z)=1+z\sum_{j=1}^kl_{j,0}\,\Phi_{j-1}(z)~~~~~~,~~~~~~
\frac1{z'^{k-1}}\,\Phi_{k-1}(z')=1+\frac1{z'}\,\sum_{j=1}^{k-1}p_{j,0}\,
\Lambda_{j-1}(z')
\label{mixediterate}\eeq
Substituting (\ref{mixediterate}) into (\ref{mixeddiff}) and comparing the
resulting equation with itself under the interchange of arguments
$z\leftrightarrow z'$, we arrive after some algebra at
\bea
\left(z-z'\right)\,\sum_{n=0}^{N-1}\frac{\Phi_n(z)\,\Lambda_n(z')}{h_n}
&=&\frac{z'^{N-1}\,\Lambda_{N-1}(z')\,\Phi_N(z)-z^{N-1}\,\Lambda_{N-1}(z)
\,\Phi_N(z')}{h_{N-1}\,z'^{N-1}}\nn\\& &+\,\frac{1-h_0}{h_0}\,\left(z-z'\right)
\label{CDgenxy}\eea
The expression (\ref{CDgenxy}) is valid for any pair of complex numbers $z\neq
z'$. We can take the $z\to z'$ limit of (\ref{CDgenxy}) using l'Hospital's rule
to get
\beq
\sum_{n=0}^{N-1}\frac{\Phi_n(z)\,\Lambda_n(z)}{h_n}=\frac{z^{N-1}\,
\Lambda_{N-1}(z)\,\Phi_N'(z)-\Phi_N(z)\,\Bigl(z^{N-1}\,
\Lambda_{N-1}(z)\Bigr)'}{h_{N-1}
\,z^{N-1}}+\frac{1-h_0}{h_0}
\label{CDgenxx}\eeq
The identities (\ref{CDgenxy}) and (\ref{CDgenxx}) are the fermionic analogs of
the Christoffel-Darboux formula for the usual Hermitian one-matrix model
orthogonal polynomials.

The Christoffel-Darboux formula allows us to express the spectral kernel
(\ref{speckernel}) in terms of orthogonal polynomials with indices $n$ close to
the matrix dimension $N$. Using the mixed recursion relation
(\ref{mixedrecrel1}), we obtain
\bea
{\cal K}(z,z')&=&\frac{\e^{\frac N2\,\bigl(V(z)+V(z')\bigr)}}
{(z\,z')^{\frac{N+1}2}}\nn\\& &\times\,\left\{
\frac{z'^{-N+1}}{h_{N-1}p_{N,0}\,(z-z')}\,\Bigl[z\,\Phi_{N-1}(z)
\,\Phi_N(z')-z'\,\Phi_{N-1}(z')\,\Phi_N(z)\Bigr]
+\frac{1-h_0}{h_0}\right\}\nn\\& &
\label{speckernelN}\eea
It follows that the fermionic matrix model is completely determined by the
single set of $\Phi$ polynomials, as expected because of the one-matrix nature
of the model and the heuristic identification $U\sim\ps2$. In particular, for
the
spectral density we find
\bea
\rho(z)&\equiv&\frac1N\,{\cal K}(z,z)\nn\\&=&\frac{\e^{NV(z)}}{Nz^N}\,\left\{
\frac1{h_{N-1} p_{N,0}\,z^N}\,\left[z \left( \Phi_N(z)\,\Phi_{N-1}'(z)-
\Phi_N'(z)\,\Phi_{N-1}(z) \right)+\Phi_N(z)\,\Phi_{N-1}(z) \,\right] \nn
\right.\\&& \left. +\,\frac{1-h_0}{h_0}\right\}
\label{specdensity}\eea

The formula (\ref{specdensity}) is particularly useful for determining the
spectral density, and hence all correlators, of the fermionic matrix model in
the large $N$ limit. In that case, factorization and symmetry imply that all
connected correlation functions vanish and the large $N$ limit of the model is
completely characterized by the set of correlators
\beq
\left\la\frac1N\,\tr\left(\ps2\right)^p\right\ra_{\rm F}=\int dz~\rho(z)\,z^p
\label{corrsrho}\eeq
where the integral goes over the support of the function $\rho(z)$ in the
complex plane. Note that the spectral density is defined {\it a priori} in the
adjoint fermion matrix model by the formula (\ref{jointprob}) \cite{mss,ss1}.
In the present formalism it has a natural interpretation as the probability
density of {\it eigenvalues} in the corresponding unitary one-matrix model. As
such, it is generally supported on the unit circle. However, as discussed in
the previous subsection, the restriction of the integration contour to the unit
circle in the large $N$ limit is not necessary, and the spectral density can be
supported generically in the complex plane. These facts explain the general
properties of the spectral densities of fermionic matrix models
\cite{acm,mss,ss1}, for instance how the pole generated by the Penner
interaction requires the support contour of $\rho(z)$ to be adjusted so as to
avoid the origin of the complex plane and why the support endpoints are
complex-valued.

\appendix{Solution of the Gaussian Model}

In this appendix we will illustrate how the formalism of section 3 works in a
specific example. We shall consider the example of the Gaussian potential
\beq
V(z)=z
\label{gausspot}\eeq
for which everything can be obtained explicitly. In this case the recursion
relations of section 3.2 are easily solved to give the coefficients
\bea
R_n&=&\frac{n}{n+N}\nn\\P_n&=&\frac{n}{N}\nn\\Q_{n}&=&- \left( 1 +
\frac{n+1}{N} \right)
\label{gausscoeffs}\eea
and the normalization constants are
\beq
h_n=2\pi i\,N^N\,\frac{n!}{(n+N)!}
\label{hngauss}\eeq
Substituting (\ref{gausscoeffs}) into the three-term recursion relation
(\ref{3termPhi}), we obtain
\beq
\Phi_n (z) = \left( z + \frac{ n - 1+N}{N} \right) \Phi_{n-1}(z) - \frac{(n-1)
z}{N}\,\Phi_{n-2}(z)
\label{gaussianrecrel}\eeq

To solve the recurrence relation (\ref{gaussianrecrel}), we introduce the
generating function which is defined as the formal power series
\beq
\Xi_N(z;s)=\sum_{n=0}^\infty \Phi_n(z)~s^n
\label{genfnphi}\eeq
in a variable $s$, with the boundary condition $\Xi_N(z;0)=1$. The recursion
relation (\ref{gaussianrecrel}) is then equivalent to a first order
inhomogeneous linear differential equation for the function $\Xi_N$,
\beq
\left[ \frac{s^2}{N}\,\Bigl( 1- s z\Bigr)\,\frac\partial{\partial
s} + s \left( z + 1 -\frac{s z}{N} \right) - 1 \right]\Xi_N(z;s)=-1
\label{genfndiffeq}\eeq
Integrating (\ref{genfndiffeq}) we find that the generating function
(\ref{genfnphi}) is given by
\beq
\Xi_N(z;s) = \frac{1}{s}\,\sum_{k=0}^\infty \frac{(N+k -1)!}{(N-1)!}\,
\left( \frac{1}{N} \right)^k\,\left( \frac{s}{1-s z } \right)^{k+1}
\label{genfnexpl} \eeq
Expanding the function (\ref{genfnexpl}) as a power series in $s$ and equating
its coefficient of $s^n$ with that of the definition (\ref{genfnphi}), we
arrive at an explicit form for the polynomials $\Phi_n(z)$,
\beq
\Phi_n(z)=\sum_{k=0}^n{n\choose k}\,
\frac{( N+ k -1)!}{(N-1)!}\,\left( \frac{1}{N} \right)^k\,z^{n-k}
\label{phinexpl}\eeq
which can be expressed as
\beq
\Phi_n(z)= N^N\,z^{N+n}\,U[ N, n+1 + N; N z]
\label{Phinhyper}\eeq
where
\beq
U[a,c;x]=\frac\pi{\sin\pi c}\,\left(\frac{{}_1F_1[a,c;x]}{\Gamma(a-c+1)
\Gamma(c)}+\frac{x^{1-c}~{}_1F_1[a-c+1,2-c;x]}{\Gamma(a)\Gamma(2-c)}\right)
\label{conflhyper}\eeq
is a confluent hypergeometric function. The partition function and observables
of the Gaussian fermionic one-matrix model are therefore completely determined
by a system of confluent hypergeometric polynomials.

\end{document}